\documentclass[twocolumn,superscriptaddress,amsmath,amssymb,floatfix,aps,pre,showpacs]{revtex4}
\usepackage{amsmath}    

\usepackage{graphicx}   
\usepackage{verbatim}   
\usepackage{color}      
\usepackage{subfigure}  
\usepackage{hyperref}   

\usepackage{amsmath}    
\usepackage{graphicx}   
\usepackage{verbatim}   
\usepackage{color}      
\usepackage{subfigure}  
\usepackage{hyphenat}
\usepackage[normalem]{ulem}
\usepackage{array}

\usepackage{color}      
\usepackage{soul}

\newcommand{\fref}[1]{Fig.~\ref{fig:#1}}

\newcommand{\frefstwo}[2]{Figs.~\ref{fig:#1}~and~\ref{fig:#2}}
\newcommand{\flabel}[1]{\label{fig:#1}}
\newcommand{\eref}[1]{Eq.~\ref{eqn:#1}}

\newcommand{\erefstwo}[2]{Eqs.~\ref{eqn:#1}~and~\ref{eqn:#2}}

\newcommand{\erefsrange}[2]{Eqs.~\ref{eqn:#1}-\ref{eqn:#2}}
\newcommand{\elabel}[1]{\label{eqn:#1}}

\newcommand{\avg}[1]{\langle #1\rangle}

\newcommand{\epsP}{\epsilon_+}
\newcommand{\epsM}{\epsilon_-}

\begin{document}
\bibliographystyle{prsty} 

\title{Optimal entrainment of circadian clocks in the presence of noise}
\author{Michele Monti} \affiliation{FOM Institute AMOLF,
  Science Park 104, 1098 XE Amsterdam, The Netherlands}
\author{David K. Lubensky}\affiliation{Department of Physics,
    University of Michigan, Ann Arbor, MI 48109-1040}
 \author{Pieter
  Rein ten Wolde} \affiliation{FOM Institute AMOLF, Science Park 104,
  1098 XE Amsterdam, The Netherlands}

\begin{abstract}
  Circadian clocks are biochemical oscillators that allow organisms to
  estimate the time of the day.  These oscillators are inherently
  noisy due to the discrete nature of the reactants and the stochastic
  character of their interactions.  To keep these oscillators in sync
  with the daily day-night rhythm in the presence of noise, circadian
  clocks must be coupled to the dark-light cycle. In this manuscript,
  we study the entrainment of phase oscillators as a function of the
  intrinsic noise in the system. Using stochastic simulations, we
  compute the optimal coupling strength, intrinsic frequency and shape
  of the phase-response curve, that maximize the mutual information
  between the phase of the clock and time. We show that the optimal
  coupling strength and intrinsic frequency increase with the noise,
  but that the shape of the phase-response curve varies
  non-monotonically with the noise: in the low-noise regime, it
  features a deadzone that increases in width as the noise increases,
  while in the high-noise regime, the width decreases with the
  noise. These results arise from a trade-off between maximizing
  stability---noise suppression---and maximizing linearity of the
  input-output, {\it i.e.} time-phase, relation. We also show that
  three analytic approximations---the linear-noise approximation, the
  phase-averaging method, and linear-response theory---accurately
  describe different regimes of the coupling strength and the noise.
\end{abstract}

\maketitle
\section{Introduction}
Many organisms possess a circadian clock to anticipate the changes
between day and night.  Circadian clocks are biochemical oscillators
that can tick without any external driving with an intrinsic,
free-running period of about 24 hrs. In uni-cellular organisms these
oscillations are formed by chemical reactions and physical
interactions between molecules inside the cell, while in
multi-cellular organisms these oscillations are typically shaped by a
combination of intra- and inter-cellular interactions, which are,
however, both mediated by molecular interactions. Due to the
discreteness of molecules and the stochastic nature of chemical and
physical interactions, circadian oscillations are inherently
stochastic, which means that they have an intrinsic tendency to run
out of phase with the day-night cycle. To keep the circadian
oscillations in phase with the day-night rhythm, the oscillations must
be coupled to daily cues from the environment, such as daily changes
in light-intensity or temperature. This coupling makes it possible to
lock the clock to, i.e. synchronize with, the daily rhythm.  However,
how the circadian clock should be coupled to entrainment cues is a
question that is still wide open. It is neither clear what the natural
performance measure for entrainment is, nor is it fully understood how
this depends on the strength and form of the coupling, the
characteristics of the entrainment signal, and the properties of the
clock.

The function that is most commonly used to describe the coupling of
the clock to the entrainment signal is called the phase-response curve
\cite{Pikovsky2003}. It gives the shift of the phase of the clock as
induced by a perturbation (a small change in, e.g., light intensity),
as a function of the phase at which the perturbation was given. The
phase-response curve has been measured for a wide variety of
organisms, ranging from cyanobacteria, to fungi, plants, flies, and
mammals \cite{Pfeuty2011}. Interestingly, these phase-response curves
share a number of characteristic features: they typically consist of a
positive and a negative lobe, and often possess a deadzone of no
coupling during the subjective day (see \fref{cartoon}). Yet, the width of the deadzone can
vary significantly, and also the negative and positive lobe are not
always equal in magnitude.

These observations naturally raise the question of what the best shape is for a phase-response curve. To answer this, a measure that quantifies the performance of the system is
needed. Several measures have been put forward. A key characteristic
of any locking scheme is the Arnold Tongue \cite{Pikovsky2003}, which
describes the range of system parameters over which the deterministic
system is locked to the driving signal. In general, this range tends
to increase with the strength of the driving signal, and one
performance measure that has been presented is how the range -- the
width of the Arnold Tongue -- increases with the magnitude of the
driving; this derivative has been called the ``entrainability'' of the
clock \cite{Hasegawa:2013es,Hasegawa:2014kf}. Another hallmark of any
stochastic system is its robustness against noise, and, in general,
the stability of an entrained clock depends not only on its intrinsic
noise, but also on the strength and shape of the coupling function;
one way to quantify clock stability is the so-called ``regularity'',
which is defined as the variance of the clock period
\cite{Hasegawa:2013es,Hasegawa:2014kf}. Another important property of
any locked system, is its sensitivity to fluctuations in the driving
signal. To quantify this, Pfeuty {\em et al.} have defined two
sensitivity measures, one that describes the change in the phase
difference between the signal and the clock due to a change in the
input, and another that quantifies the change in the stability of the
fixed point (the slope of the phase-response curve) in response to a
change in the input signal \cite{Pfeuty2011}.

These performance measures make it possible to make predictions on the
optimal shape of the phase-response curve. Pfeuty {\em et al.} argued
that the shape of the phase-responce curve is determined by the
requirement that the clock should respond to changes in light
intensity that are informative on the day-night rhythm, namely
light-intensitiy changes during dawn and dusk, but should ignore
uninformative fluctuations in light intensity during the day, arising,
e.g., from clouds \cite{Pfeuty2011}. This naturally gives rise to a
deadzone in the phase-response curve, which allows the clock to ignore
the input fluctuations during the day. Hasegawa and Arita argued
that the shape of the phase-resopnse curve is determined by a
trade-off between regularity (stability) and entrainability
\cite{Hasegawa:2013es,Hasegawa:2014kf}. Entrainability requires not
only changes in light intensity, but also that a change in the copy
number $n_i$ of a component $i$, as induced by the changing light
signal, leads to a change in the phase $\phi$ of the clock: the gain
$d\phi/dn_i$ should be large. However, a higher gain also means that
the evolution of the phase becomes more susceptible to noise in
$n_i$. Maximizing entrainibility for a given total noise strength
integrated over 24 hrs then yields a phase-response curve with a
deadzone: During the day, when informative variations in light
intensity are low, a high gain will not significantly
enhance entrainability but will increase the integrated noise, implying that the gain should be as low as possible during the middle of the day.

In this manuscript, we introduce another measure to quantify the
performance of the system, the mutual information
\cite{Shannon1948}. The mutual information quantifies the number of
signals that can be transmitted uniquely through a communication
channel. As such it is arguably the most powerful measure for
quantifying information transmission, and in recent years the mutual
information has indeed been used increasingly to quantify the quality
of information transmission in cellular signaling systems
\cite{Ziv2007,Tostevin2009,Mehta2009,Tkacik:2009ta,tostevin10,DeRonde2010,Tkacik2010,Walczak:2010cv,DeRonde2011,Cheong:2011jp,deRonde:2012fs,Dubuis2013,Bowsher:2013jh,Selimkhanov:2014gd,DeRonde:2014fq,Govern:2014ez,Sokolowski:2015km,Becker2015,
  Monti2016}.  In the context studied here, the central idea is
that the cell needs to infer from a variable of the clock, e.g. its
phase $\phi$, the time of the day $t$. The mutual information then
makes it possible to quantify the number of distinct time points that
can be inferred uniquely from the phase of the clock. Importantly, how many
time states can be inferred reliably, depends not only on the noise in
the system, but also on the shape of the input-output curve,
$\overline{\phi}(t)$, i.e. the average phase $\overline{\phi}(t)$ at time $t$.

We study how the mutual information between the clock phase and the time depends on the shape and magnitude of the phase response curve in the presence of intrinsic noise in the system; we
thus do not consider fluctuations in the input signal. The clock is
modeled as a phase oscillator and the phase-resopnse curve is
described via a piecewise linear function (see \fref{cartoon}), which
allows for optimization and analytical results. We find that for a
given amount of noise in the system there exists an optimal coupling
strength that maximizes the mutual information: Increasing the
coupling strength too much will decrease the mutual
information. However, as the noise in the system increases, the
optimal coupling strength increases. Moreover, for a given shape of
the phase-response curve featuring a deadzone, the optimal
intrinsic (free running) period of the clock is non-monotonic: as the
noise is increased, the optimal period first becomes larger than 24
hrs, but then decreases to become smaller than 24 hrs. Optimizing over
not only the coupling strength and the intrinsic period, but also over
the shape of the phase response curve, reveals that the optimal width
of the deadzone is also non-monotonic. As the noise is increased, the width
first increases, but then decreases. We show that all of these results
can be understood as a trade-off between linearity and stability. At
low noise, it is paramount to make the input-output relation $\overline{\phi}(t)$
as linear as possible, because this maximizes the mutual information;
this is enhanced by a large deadzone and weak coupling. However, for
large noise strengths, stability becomes key, which favors a small
deadzone, a stronger coupling, and a smaller intrinsic period.

In the next section, we first briefly present the Chemical Langevin
Description of a biochemical network, because this is important for
understanding not only the phase-reduction method that reduces the
system to a phase-oscillator model, but also for unstanding some
important characteristics of the mutual information. In the subsequent
section, we then introduce the mutual information. We emphasize that
the mutual information is insensitive to a coordinate transformation
and that the mutual information between all degrees of freedom of the
system (i.e. copy numbers of all components) and the input (i.e. time
$t$) is always larger than that between one degree of freedom and the
input. This means that the mutual information that we will compute
between the phase of the clock and the time will provide a firm lower
bound on the actual mutual information. We then briefly describe our
phase-oscillator model and how we model the phase-response curve.

In the results section, we first present the results of stochastic
simulations of our phase-oscillator model. By performing very
extensive simulations we find the system parameters that maximize the
mutual information, and by explicitly computing the linearity and
stability as a function of parameters, we show that the optimal design
as a function of the noise arises from the trade-off
mentioned above between linearity and stability.

Finally, we present and apply three different analytic approximations (or ``theories''), and show that
each recapitulates the simulations in a different parameter regime.  The linear-noise approximation accurately describes the regime
of low  noise and strong coupling.  The phase-averaging method \cite{Pikovsky2003} captures the
regime of low noise and weak coupling.   Finally, the linear-response
theory accurately describes the mutual information in the regime of
high noise and weak coupling.  Whereas the first two approximations are valid in the vicinity of the optimal coupling for an appropriate range of noise strengths, the third turns out to hold only far from optimality.

\section{Model}
\subsection{Chemical Langevin Description}
We consider a self-sustained oscillator of $M$ components with copy
numbers $n_1, n_2, \dots, n_M$, denoted by the vector ${\bf n}$. Its
dynamics is given by
\begin{align}
\frac{d{\bf n}}{dt} &={\bf A}({\bf n}),
\elabel{dndt}
\end{align}
where ${\bf A}({\bf n})$ is determined by the propensity functions of
the chemical reactions that constitute the network. The limit cycle of
the free-running oscillator is the stable periodic solution of this
equation, ${\bf n}(t) = {\bf n}(t+T_0)$, where $T_0$ is the intrinsic
period of the oscillator.

Due to the stochasticity of the chemical reactions and the
discreteness of the molecules, the evolution of the network is
stochastic. When the copy numbers are sufficiently large such that there
exists a macroscopic time interval $dt$ during which the propensity
functions remain constant and the Poissonian distribution of reaction
events can be approximated as a Gaussian,  then
the dynamics is described by the chemical Langevin equation \cite{Gillespie:2000vv},
\begin{align}
\frac{d{\bf n}}{dt} &={\bf A}({\bf n}) + {\underline \eta}({\bf n}),
\elabel{dndtnoise}
\end{align}
where the vector $\underline{\eta}(t)$ describes the Gaussian white noise,
characterized by the noise matrix with elements  $\langle \eta_i({\bf
  n }(t))
  \eta_j ({\bf n}(t^\prime))\rangle = D_{ij}({\bf n}) \delta (t-t^\prime)$.
 
  A clock is only a useful timing device if it has a stable and
  precise phase relationship with the daily rhythm. Biochemical noise
  tends to disrupt this relationship. To keep the clock in sync with
  the day-night rhythm in the presence of noise, the clock must be coupled
  to the light signal:
\begin{align}
\frac{d{\bf n}}{dt} &={\bf A}({\bf n}) + \epsilon {\bf p}({\bf n},t) + {\underline \eta}({\bf n}).
\elabel{dndtForceNoise}
\end{align}
Here ${\bf p}({\bf n},t)$ describes the coupling to the light signal and
$\epsilon$ the strength of the coupling. The coupling force ${\bf
  p}({\bf n},t) = {\bf p}({\bf n},t+T)$ has a
period $T$ and frequency $\omega=2\pi / T$, which in general is different from the intrinsic period
$T_0$ and intrinsic frequency $\omega_0=2\pi / T_0$, respectively, of the free-running oscillator. In this manuscript, we will assume that the light signal is
deterministic. We thus only consider the biochemical noise in the clock.

\subsection{Mutual information}
The organism needs to infer the time $t$ from the concentrations of
the clock components. This inference will be imprecise, because of the
noise in the clock. We will quantify the accuracy of information
transmission via the mutual information, which is a measure for how
many distinct time states can be resolved from the concentrations of
the clock components \cite{Shannon1948}.

The mutual information $I({\bf n}; t) = I(\{n_1, \dots, n_M\};t)$
between the copy numbers of all components and the time is given by 
\begin{align}
I({\bf n};t) &= \int d{\bf n} \int dt P({\bf n};t) \log_2
\frac{P({\bf n};t)}{P({\bf n}) P(t)},
\end{align}
where $P({\bf n};t)$ is the probability that copy numbers ${\bf n}$ are found at time $t$. $I({\bf n};t)$ measures the reduction in uncertainty about $t$ upon measuring
$\{n_1,\dots,n_M\}$, or vice versa. The quantity is indeed symmetric in ${\bf
  n}$ and $t$:
\begin{align}
I{\bf n};t)&=H(t) - \avg{H(t|{\bf n})}_{\bf n}\\
&=H({\bf n}) - \avg{H({\bf n}|t)}_t,
\end{align}
where $H({\bf a}) = -\int d{\bf a} P({\bf a})  \log_2 P({\bf a})$,
with $P({\bf a})$ the
probability distribution of ${\bf a}$, is the entropy of ${\bf a}$;
$H({\bf a}|{\bf b}) = - \int d{\bf a} P({\bf a}|{\bf b}) \log_2 P({\bf
  a}|{\bf b})$ is the
information entropy of ${\bf a}$ given ${\bf b}$, with $P({\bf a}|{\bf
  b})$ the conditional probability distribution of ${\bf a}$ given
${\bf b}$; $\avg{f(c)}_c$ denotes an average of $f(c)$ over the
distribution $P(c)$.

A key point worthy of note is  that the mutual information is
invariant under a coordinate transformation, which allows us to put a
firm lower bound on the mutual information between time and the clock
components.
 Specifically, we can first make a non-linear
transformation from ${\bf n}$ to some other set of
variables ${\bf x}$, of which two components are the amplitude $R$ of
the clock and its phase $\phi$. Because the mutual information is
invariant under this transformation,
\begin{align}
  I({ \bf n},t)=I({\bf
  x},t).
\elabel{Icrdtrnsf}
\end{align}
 Secondly, if the time is inferred not from all the
components of ${\bf x}$, but rather from $R$ and $\phi$, then, in
general
\begin{align}
I(R,\phi;t) \leq I({\bf x};t).
\end{align}
By combining this expression with \eref{Icrdtrnsf}, we find that
\begin{align}
I({\bf n};t) \geq I(R,\phi;t)
\end{align}
Hence, once we have defined a mapping between ${\bf n}$ and ${\bf x}$
and hence
$(R,\phi)$, the mutual information $I(R,\phi;t)$ between the
combination of the amplitude
and phase of the clock $(R,\phi)$ and time $t$, puts a lower bound on
the mutual information $I({\bf n};t)$. A weaker lower bound is provided
by the mutual information between the phase of the clock and time:
\begin{align}
I({\bf n};t) \geq I(R,\phi;t) \geq I(\phi;t).
\elabel{Ibound}
\end{align}
However, we expect this bound to be rather tight, since a reasonable,
natural, mapping between ${\bf n}$ and $(R,\phi)$ should put the
information on time in the phase of the clock.

\subsection{Phase oscillator}
The bound of \eref{Ibound} makes it natural to develop a description
of the clock in terms of the phase.  Here, we review the derivation of such a description, largely following the standard arguments in \cite{Pikovsky2003}, but paying special attention to the appropriate form of the effective noise on the phase variable.  In the absence of any
coupling and noise, the temporal evolution of the phase is given by
\begin{align}
\frac{d\phi({\bf n})}{dt} = \omega_0,
\end{align}
where $\omega_0 = 2\pi / T_0$ is the intrinsic frequency of the
clock, with $T_0$ the intrinsic period. As the phase is a smooth
function of ${\bf n}$, the evolution of $\phi$ is also given by
\begin{align}
\frac{d\phi({\bf n})}{dt} = \sum_i \frac{\partial \phi}{\partial n_i}
  \frac{dn_i}{dt}.
\elabel{dphidtdndt}
\end{align}
Combining the above two equations with \eref{dndt} yields the
following expression for the intrinsic frequency
\begin{align}
\omega_0 = \sum_i \frac{\partial \phi}{\partial n_i} A_i ({\bf n}).
\elabel{omega_0}
\end{align}
This equation {\em defines} a mapping $\phi({\bf n})$. This mapping is
defined such that for each point ${\bf n}$ in state space, the
time derivative $d\phi({\bf n})/dt = d\phi/dt$ of the phase is constant and
equal to $\omega_0$. The surfaces of constant $\phi({\bf n})$, defined
according to this mapping, are called isochrones.  

In the presence of noise, the phase dynamics is,
combining \erefstwo{dndtnoise}{dphidtdndt},
\begin{align}
\frac{d\phi({\bf n})}{dt} &= \sum_i \frac{\partial \phi}{\partial n_i}\left[
  A_i ({\bf n}) + \eta_i ({\bf n})\right],\\
&= \omega_0 + \xi({\bf n}), \elabel{phi-langevin}
\end{align}
which yields for the noise on the phase variable
\begin{align}
\xi({\bf n}) =  \sum_i \frac{\partial \phi}{\partial n_i}\eta_i({\bf n}).
\end{align}
In general, the variance of $\xi$ thus depends on all of the state variables ${\bf n}$, not just on the phase $\phi$, and \eref{phi-langevin} does not give a closed description in terms only of $\phi$.
 However, when the deviations from the limit cycle are small
compared to the scale over which the noise strength changes as a function of distance from the limit cycle, we can estimate the noise by evaluating it at the
limit cycle, ${\bf n}_0$:
\begin{align}
\xi(\phi) =  \sum_i \frac{\partial \phi({\bf n}_0)}{\partial n_i}\eta_i({\bf n}_0).
\end{align}
with Gaussian white noise statistics
\begin{align}
\avg{\xi(\phi(t)) \xi(\phi(t^\prime))} &=\sum_{i,j} \frac{\partial
  \phi}{\partial n_i} \frac{\partial \phi}{\partial n_j} D_{ij} ({\bf n}_0) \delta
(t-t^\prime),\\
&\equiv 2 D (\phi) \delta (t-t^\prime).
\end{align}


When the system is coupled to light, the phase evolution becomes, from \erefstwo{dndtForceNoise}{dphidtdndt},
\begin{align}
\frac{d\phi({\bf n})}{dt} &=  \sum_i \frac{\partial \phi}{\partial n_i}\left[
  A_i ({\bf n}) +  \epsilon 
  p_i({\bf n},t) + \eta_i ({\bf n})\right].
\end{align}
The force depends explicitly on time. This impedes a unique definition
of the isochrones $\phi({\bf n})$, because how the phase evolves at a
particular point in phase space depends not only on ${\bf n}$ but also
on $t$. Of course,  one could still adopt the mapping of the free running
system, in which case the evolution of the phase is given by
\begin{align}
\frac{d\phi({\bf n})}{dt}&= \omega_0 + \epsilon\sum_i \frac{\partial
  \phi}{\partial n_i} p_i({\bf n},t) + \xi(\phi).
\end{align}
The problem is that, because along the surface $\phi({\bf n})$ the
light-coupling term is not constant, $d\phi({\bf n})/dt$ will depend on
${\bf n}$. One can then not reduce the dynamics to that of a single
phase variable.
 
However, if $\epsilon$ is small
and the force only leads to small deviations from the limit cycle of
the free-running system, then one may approximate the effect of the forcing by evaluating the corresponding term at the limit cycle, ${\bf n}_0$. We then have
\begin{align}
\frac{d\phi({\bf n})}{dt} &=  \omega_0 + \epsilon\sum_i \frac{\partial
  \phi({\bf n}_0)}{\partial n_i} p_i({\bf n}_0,t) + \xi(\phi).
\end{align} 
In this case the evolution of
the phase no longer explicitly depends on ${\bf n}$:
\begin{align}
\frac{d\phi}{dt} = \omega_0 + Q(\phi,t) + \xi (\phi),
\end{align}
with 
\begin{align}
Q(\phi,t) = \epsilon \sum_i \frac{\partial
  \phi({\bf n}_0(\phi))}{\partial n_i} p_i({\bf n}_0 (\phi),t).
\end{align}

How a circadian clock responds to a given light signal $L(t)$ depends
on its phase $\phi$; it does not explicitly depend on time. The
coupling term can then be written as $Q(\phi,t) =
Z(\phi) L(t)$, where $Z(\phi)$ is the
instantaneous phase response curve, which describes how the clock
responds to the light signal as a function of its phase $\phi$.
In addition, while in general the noise strength depends on the phase, we will,
motivated by the experimental observations of Mihalecescu and Leibler
on the {\it S. elongatus} clock\cite{Mihalcescu:2004ch},
assume it is
constant. We
then finally arrive at the equation that describes the evolution of
the phase in our model:
\begin{equation}
\frac{d\phi}{dt} = \omega_0 + Z(\phi)L(t) + \xi (t),
\elabel{dphidt}
\end{equation}
with $\langle \xi(t)
\xi(t^\prime)\rangle = 2D \delta (t-t^\prime)$.

In what follows, we will study entrainment using the above equation not only when $Z(\phi)L(t)$ and $D$ are much smaller than $\omega_0$, so that the weak coupling assumptions necessary for the reduction to a phase oscillator clearly hold, but also when $Z(\phi)L(t)$ or $D$ are of order $\omega_0$ or larger.  As we discuss in more detail in Section \ref{sec-discussion}, however, this does not present any contradiction, because it is perfectly possible for the noise and the external driving to be small compared to restoring forces orthogonal to the limit cycle, so that the system always stays near the limit cycle and the phase is the only relevant variable, while simultaneously strongly perturbing motion along the limit cycle.
We also note that $\epsilon$ can be varied
independently of the noise strength. What is perhaps less obvious is
whether $Z(\phi)$ and $D$ can be varied independently. When the
  size of the system, e.g. the volume of the living cell, is changed,
  as was done for {\it Bacillus subtilus} \cite{Liu:2007dm}, then
  the noise strength $D$ will change, but the coupling strength
  $Z(\phi)$ will, to first order, not change because the
  concentrations remain constant. Moreover, typically the system is
  coupled to light only via a relatively small number of reactions, while the noise is
  determined by all reactions.  Also in this case, it seems natural to
  assume that $Z(\phi)$ and $D$ can be varied independently. We note
  that the arguments of Hasegawa and Arita do not contradict our
  arguments that $Z(\phi)$ and $D$ can be varied independently: the fact that changing the gain $\partial \phi/\partial
  n_i$ affects both the coupling to light (entrainability) and the
  phase noise \cite{Hasegawa:2013es,Hasegawa:2014kf}, does not
  mean that the noise and the coupling cannot be varied independently if other parameters are changed
  (and vice versa). We thus imagine that $p_i ({\bf n})$ can be tuned
(by evolution) independently of the $D_{ij}({\bf n})$. We do not
change the mapping $\phi({\bf n})$, determined by the properties of
the uncoupled system.

\subsection{The system}
We will approximate $Z(\phi)$ and $L(t)$ as step functions, shown in
 \fref{cartoon}.  This makes it possible to analytically obtain
the Arnold tongue, i.e. the range of parameters for which the
deterministic system locks to the
day-night rhythm in the absence of noise. The light-dark function $L(t)$ is unity for
$0<t<T/2$ and zero for $T/2<t<T$.  The shape of the instantaneous
phase response curve $Z(\phi)$ is inspired by experimentally
characterized response curves, featuring a positive lobe, a dead-zone
in which $Z(\phi)$ is essentially zero, a negative lobe, followed by a
positve lobe again \cite{Pfeuty2011}. It is characterized by five
variables, the coupling strengths $\epsilon_+$ and $\epsilon_-$, and
the phases $\phi_1, \phi_2, \phi_3$:
\begin{equation}\label{IPRC}
Z(\phi) = \left\{
\begin{array}{cc}
\epsilon_+ & 0 < \phi < \phi_1 \\
0 & 	\phi_1 < \phi < \phi_2\\
-\epsilon_- & \phi_2 <\phi < \phi_3\\
\epsilon_+ & \phi_3 <\phi < 2\pi 
\end{array}
\right.
\end{equation}
where $\epsilon_+ $ and $\epsilon_-$ are greater than 0.  With these 5
variables, a wide range of experimentally characterized phase response
curves can be described.

\begin{figure*}[t]
\centering
\includegraphics[scale=0.6]{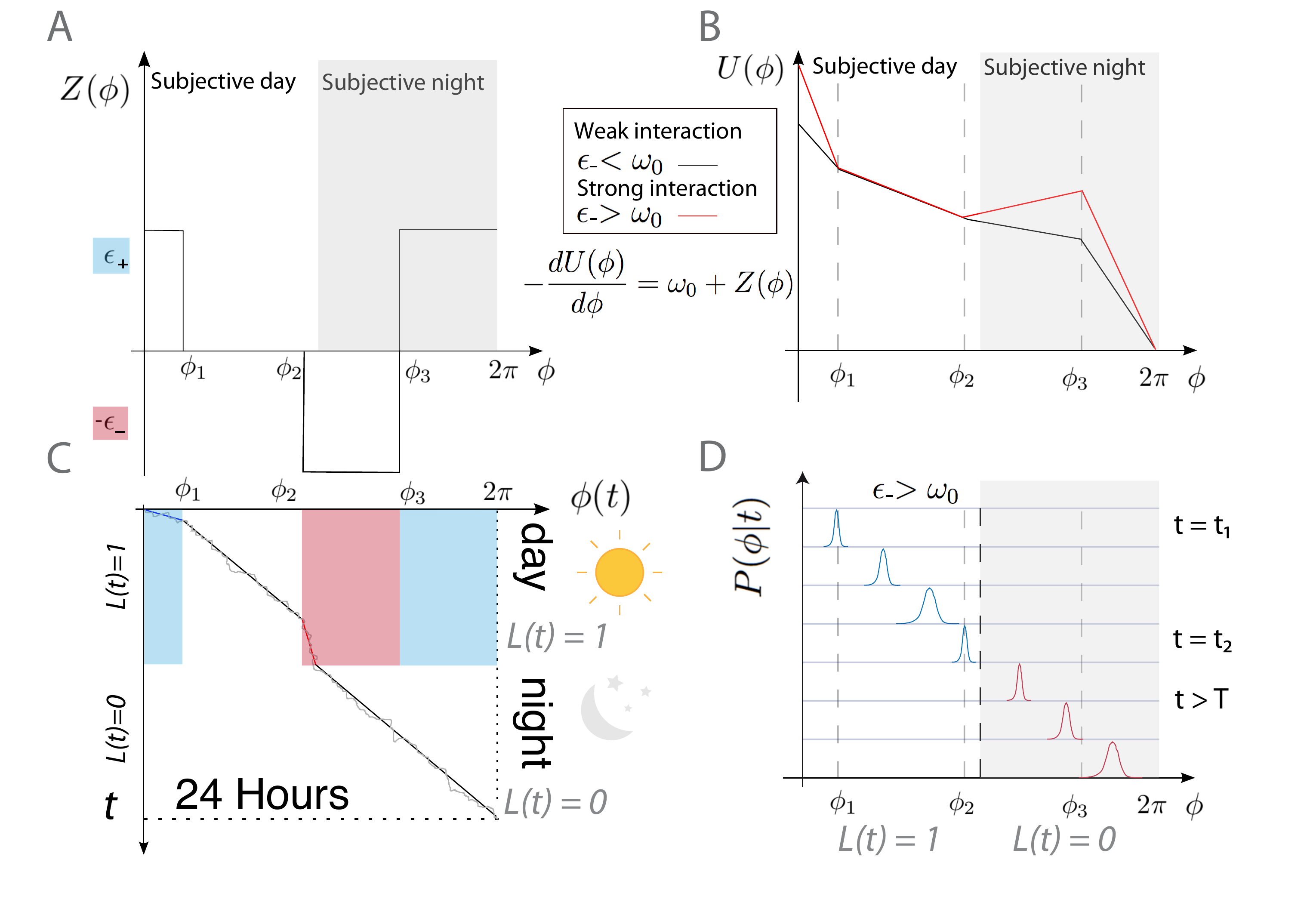}
\caption{Cartoon of the system. (A) The instantaneous phase response
  curve $Z(\phi)$, characterized by the 5 parameters $\epsilon_+,
  \epsilon_-$ and $\phi_1, \phi_2, \phi_3$. The driving signal is
  given by $L(t) = 1$ during the day and $L(t)=0$ during the night. (B) The phase evolution of
  the system, $d\phi/dt$, can be interpreted as that of a particle in a
  potential $U(\phi)$, with a force $-dU(\phi)/d\phi = 
  \omega_0 + Z(\phi) L(t)$. Note that the particle
  only experiences a force during the day, when $L(t)=1$, and not
  during the night, when $L(t)=0$. (C) The phase evolution of the
  system, in the limit of small noise. During the night the
  deterministic system
  always evolves with its intrinsic frequency $\omega_0$. During the
  day, it evolves with its intrinsic frequency $\omega_0$ when the phase is
  between $\phi_1$ and $\phi_2$; between $\phi_3 - 2\pi$ and $\phi_1$,
  the system is ``pushed'', moving with a frequency $\omega_0+\epsilon_+$,
  while between $\phi_2$ and $\phi_3$ it is slowed down, moving at
  frequency $\omega_0-\epsilon_-$. (D) Illustration of how $P(\phi)$
  evolves in time, in the regime of strong coupling. At dawn, the
  system is pushed, narrowing the distribution; during the deadzone
  in which $Z(\phi)=0$,
  the distribution tends to widen; near dusk, the system is slowed
  down, narrowing the distribution; during the night, the system
evolves freely, widening the distribution again. }
\flabel{cartoon}
\end{figure*}

\section{Results}

\subsection{Arnold Tongue of the deterministic system}
\label{sec:AT}
Motivated by the observation that circadian clocks typically lock 1:1
to the day-night rhythm, we will focus on this locking scenario,
although we will also see that this system can exhibit higher order
locking, especially when the intrinsic period of the clock deviates
markedly from that of the day-night rhythm. To derive the Arnold
tongue,  we first note that when the
clock is locked to the light-dark cycle, it will have a characteristic
phase $\phi_s$ at the beginning of the light-dark cycle, $t_s=0$. In
the case of 1:1 locking, the
phase of the clock will then cross phase $\phi_1$ at time $t_1$,
$\phi_2$ at time $t_2$, and $\phi_3$ at time $t_3$. To obtain the
Arnold tongue, we have to recognize that there are in total 12
possible locking scenarios: 3 for $\phi_s$ and 4 for
$t_1,t_2,t_3$. The scenarios for $\phi_s$ are: 1: $\phi_3 - 2 \pi <
\phi_s < \phi_1$; 2: $\phi_1 < \phi_s < \phi_2$; 3: $\phi_2 < \phi_s <
\phi_3$.  The 4 scenarios for $t_1,t_2, t_3$ are defined by where
$T/2$ falls with respect to these times: 1: $T/2 < t_1 < t_2 < t_3$;
2: $t_1 < T / 2 < t_2 < t_3$; 3: $t_1 < t_2 < T/2 < t_3$; 4: $t_1 <
t_2 < t_3 < T / 2$. For each of these 12 scenarios, we can
analytically determine $\phi_s$ and $t_1, t_2, t_3$, which then
uniquely specify $\phi(t)$. The 4 unknowns, $\phi_s,
t_1, t_2, t_3$, give each an inequality for $T$, and the range of $T$
that satisfies all 4 inequalities determines the width of the Arnold
tongue. For each of the 12 scenarios for the given $\epsilon_+,
\epsilon_-$, we have an Arnold tongue, and those 12 tongues together
give ``the'' Arnold tongue for those values of $\epsilon_+,
\epsilon_-$. We now derive the tongue for scenario 1, which is also
the most important one, as we will see: in this regime, the mutual
information between time and the phase of the lcock is the largest.

Scenario 1 is characterized by: $\phi_3 - 2\pi < \phi_s < \phi_1$;
$0<t_1 < t_2 < T/2 < t_3$. The solution depends on whether $\epsilon_-$ is
larger or smaller than $\omega_0$. If $\epsilon_-< \omega_0$, then the
deterministic system locks 1:1 to the driving signal when
\begin{align}
&\phi_s + (\epsilon_+ + \omega_0) t_1 + \omega_0 (t_2 - t_1)\nonumber\\
&+ (-\epsilon_- + \omega_0) (T/2 - t_2) + \omega_0 T / 2 = \phi_s + 2
\pi.
\end{align}
To solve this, we note that $\phi_1 = \phi_s + (\omega_0 + \epsilon_+)
t_1$, $\Delta \phi_{12} \equiv \phi_2 - \phi_1 = \omega_0 (t_2 -
t_1)$. The solution is
\begin{align}
t_1 &= \frac{2\pi - T(\omega_0 - \epsilon_-/2) - \epsilon_- \Delta
  \phi_{12}/\omega_0}{\epsilon_++\epsilon_-}\geq 0,\\
t_2 &=\frac{\Delta\phi_{12}}{\omega_0}+t_1 < T/2,\\
t_3 &=\frac{\Delta\phi_{23}}{\omega_0 - \epsilon_-}+t_2>T/2,\\
\phi_s &=\phi_1 - (\omega_0 + \epsilon_+) t_1 > \phi_3 - 2\pi,
\end{align}
where $\Delta \phi_{23}\equiv \phi_3 - \phi_2$. 
The above inequalities lead to the following inequalities for the
period $T$, respectively:
\begin{align}
T &\leq \frac{2\pi - \epsilon_-\Delta
  \phi_{12}/\omega_0}{\omega_0 - \epsilon_-/2},\\ 
T &> \frac{2\pi + \epsilon_+\Delta \phi_{12}
  /\omega_0}{\epsilon_+/2+\omega_0},\\
T& <\frac{2\pi + \epsilon_+\Delta \phi_{12}
  /\omega_0 + \Delta \phi_{23}(\epsilon_++\epsilon_-)/(\omega_0 - \epsilon_-)}{\epsilon_+/2+ \omega_0},\\
T&>\frac{(\Delta
  \phi_{13}-2\pi)(\epsilon_++\epsilon_-)/(\omega_0+\epsilon_+) + 2 \pi  - \epsilon_-\Delta
  \phi_{12}/\omega_0}{\omega_0 - \epsilon_-/2},
\end{align}
where $\Delta \phi_{13} \equiv \phi_3 - \phi_1 = \Delta \phi_{12}+\Delta \phi_{23}$.
The width of the Arnold tongue is given by the range of $T$ that
satisfies all inequalities. 

If $\epsilon_-> \omega_0$, then the equation to solve is:
\begin{align}
\phi_s + (\epsilon_+ + \omega_0) t_1 + \omega_0 (t_2 - t_1)+ \omega_0 T / 2 = \phi_s + 2 \pi.
\end{align}
The solution is
\begin{align}
t_1 &= \frac{2\pi - \omega_0 T / 2 - \Delta \phi_{12}}{\epsilon_+ + \omega_0}\geq 0\elabel{t1}\\
t_2 &=\frac{\Delta\phi_{12}}{\omega_0}+t_1 < T/2\elabel{t2}\\
t_3 &=\infty>T/2\\
\phi_s &=\phi_1 - (\omega_0 + \epsilon_+) t_1 > \phi_3 - 2\pi\elabel{phis}
\end{align}
The third inequality, for $t_3$ does not contribute, if the other
inequalities are satisfied. We thus have 3 inequalities:
\begin{align}
T& \leq \frac{2 (2\pi - \Delta \phi_{12})}{\omega_0}\\
T&> \frac{2 \pi  + \epsilon_+ \Delta
  \phi_{12}/\omega_0}{\epsilon_+/2 + \omega_0}\\
T&>\frac{2\Delta \phi_{23}}{\omega_0}
\end{align}

It is seen that the locking region does not depend on the absolute
values of $\phi_1, \phi_2, \phi_3$, but only on the separation between
them, leaving only two independent parameters that are related to the
phase: $\Delta \phi_{12}=\phi_2-\phi_1$ and $\Delta
\phi_{23}=\phi_3-\phi_2$; the remaining interval is given by $2\pi -
\Delta \phi_{13} = 2\pi - (\Delta \phi_{12}+\Delta
\phi_{23})$. Shifting the absolute values of $\phi_1,\phi_2,\phi_3$
only changes the definition of the phase of the clock, not the moments
of the day---$t_1,t_2,t_3$---at which $Z(\phi)$ changes. The system
thus has 5 independent parameters, 4 related to $Z(\phi)$---$\Delta
\phi_{12}, \Delta \phi_{23}, \epsP,\epsM$---and one being the
intrinsic frequency $\omega_0$.

In the appendix, we derive the Arnold Tongues for the other scenarios.
It turns out that only scenarios 1 - 4 yield stable solutions; the
solutions of the other scenarios are unstable. 

\fref{ArnoldTong} shows the Arnold Tongues for the 4 scenarios. Since
we imagine that the period of the light-day cycle is fixed while the
clock can adjust its intrinsic frequency $\omega_0$, we plot the range
of $\epsilon=\epsP=\epsM$ over which the system exhibits a stable
deterministic solution, as a function of $\omega_0/\omega$; $\Delta
\phi_{12}=\Delta \phi_{23} = \pi/2$. The different colors correspond
to the different scenarios. Clearly, the Arnold Tongues of the
respective scenarios are adjoining.  The region in the middle, around
$\omega_0 = \omega$, bounded by the blue lines, corresponds to our
natural scenario, i.e. scenario 1, discussed above. The green lines
bound the Arnold Tongue of scenario 3. This is an unnatural scenario,
because in this scenario the clock is driven backwards when the light
comes up. Moreover, for $\omega_0/\omega>2$, the system can also
exhibit higher-order locking, which is biologically irrelevant. We
will therefore focus on the regime $0.5 < \omega_0 < 2.$

\fref{ArnoldTong} shows that for $\epsilon<1$ the Arnold Tongue
exhibits the characteristic increase in its width as the coupling
strength is increased: coupling increases the range of frequencies
over which the clock can be entrained. However, for $\epsilon>1$, the
width does not change significantly; in fact, it does not change at
all when $\omega_0>\omega$. This is because a) during the day, for
$\epsilon_-=\epsilon=\epsilon_+>1$, the phase evolution comes to a halt at $\phi_3$---the
particle sits in the potential well of \fref{cartoon}B and b) during
the night the system evolves with a fixed speed $\omega_0$,
independent of $\epsilon$.

\begin{figure}[h!]
\centering
\includegraphics[width=8.5cm]{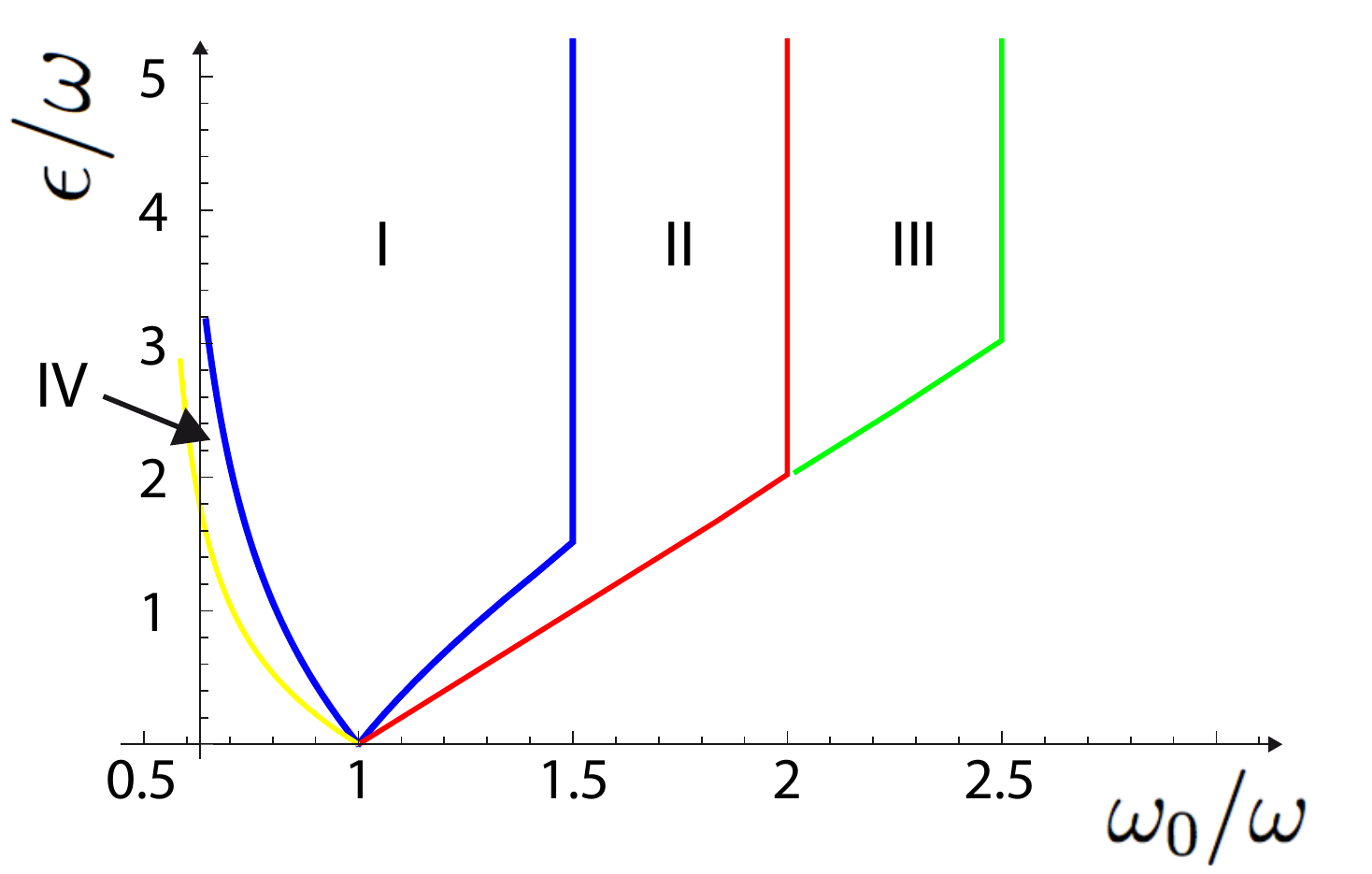}
\caption{The Arnold Tongue for 1:1 locking in the deterministic model,
  with the coupling strength $\epsP = \epsM=\epsilon$ in units of the
  (fixed) frequency of the day-night rhythm $\omega$, plotted
  as a function of the intrinsic frequency of the clock,
  $\omega_0/\omega$. The different colors correspond to the different
  scenarios that yield a stable solution. The large region around
  $\omega_0/\omega = 1$, bounded by the blue lines,
  corresponds to the Arnold Tongue of scenario 1. The adjoining region
  to the right, with the red boundaries, corresponds to scenario
  2. The green lines bound the Arnold Tongue of scenario 3, and the
  yellow lines on the far left yield the Arnold Tongue of scenario
  4. The other key parameters of $Z(\phi)$ are kept constant: $\Delta
  \phi_{12} = \Delta \phi_{23}=\pi/2$. \flabel{ArnoldTong}}
\end{figure}

\subsection{Optimal coupling strength and intrinsic frequency in
  presence of noise}
\label{sec:MutInfo}
While the Arnold Tongue shows the range of parameters over which the
deterministic system can exhibit stable 1:1 locking, it does not tell
us how reliably the time can be inferred from the phase in the
presence of noise. To address this question, we have computed the
mutual information $I(\phi;t)$ between the phase of the clock,
$\phi(t)$, and the time $t$, by performing long stochastic simulations of
the system, i.e. stochastically propagating
\eref{dphidt}. 

\fref{MutInfo}A shows a heatmap of the mutual information as a function
$\epsP=\epsM=\epsilon$ and $\omega_0/\omega$, for $\Delta
\phi_{12}=\Delta \phi_{23}=0.5\pi$ and $D = 0.1/T$. Superimposed over
the heatmap are the deterministic Arnold Tongues for scenarios 1--4, which are also shown in \fref{ArnoldTong}. It is seen that the
mutual information is highest in the region bounded by the Arnold
Tongue of 1:1 locking in scenario 1. Interestingly, however, the
figure does also show that the mutual information can be large outside
of the 1:1 locking regimes, especially when $\omega_0/\omega >
2$. This is the result of higher order locking.

\begin{figure*}[t]
\centering
\includegraphics[width=18cm]{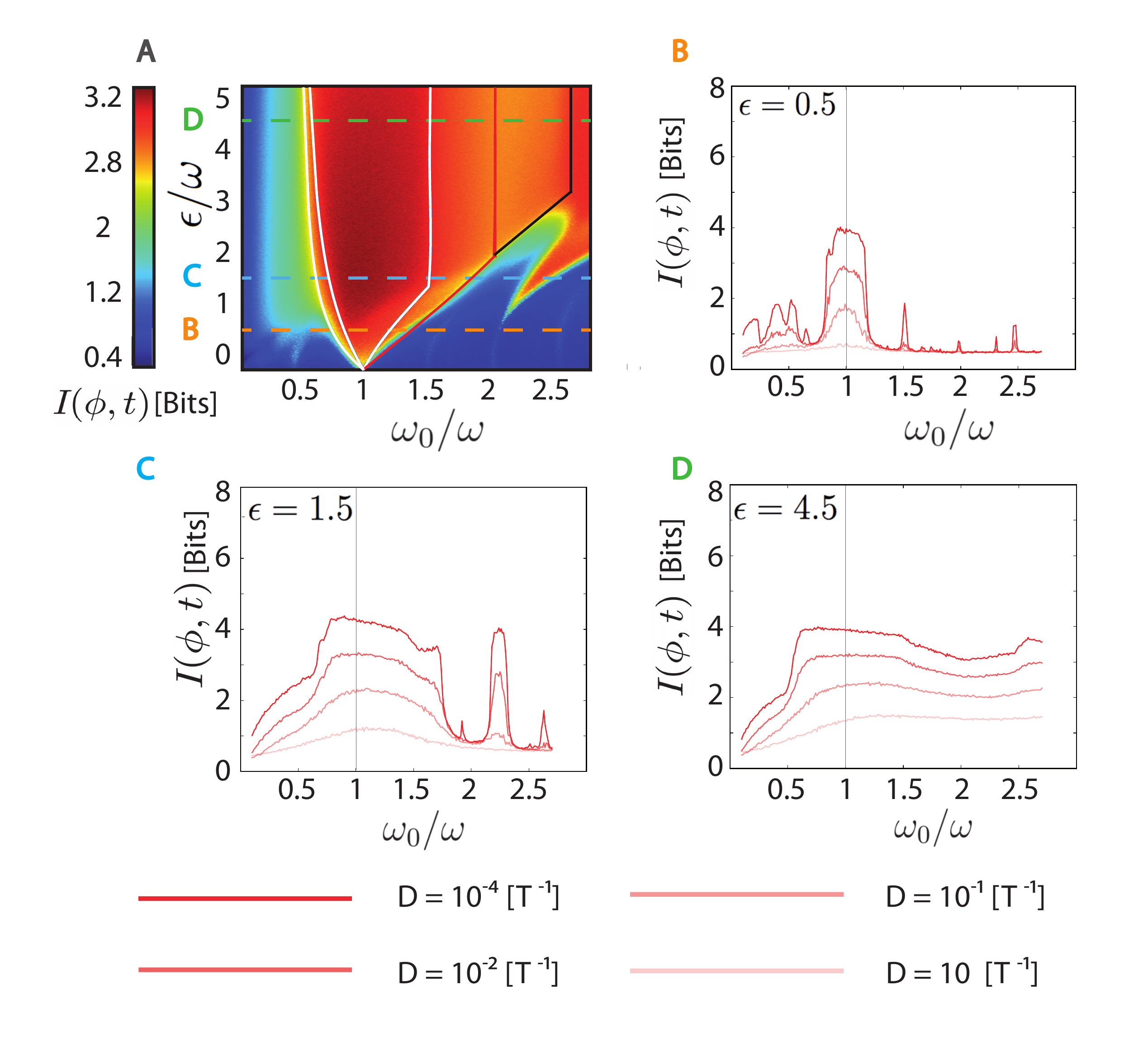}
\caption{The mutual information as a function of $\epsilon$, $D$, and
  $\omega_0$, keeping $\Delta \phi_{12}=\Delta \phi_{23}=\pi/2$. (A)
  Heatmap of the mutual information as a function of $\epsilon/\omega$
  and $\omega_0/\omega$ for $D=0.1/T$, respectively. Superimposed are the Arnold
    Tongue for 1:1 locking in scenarios 1--4. It is seen that
  the mutual information is high inside the Arnold Tongues, with the
  region corresponding to scenario 1 being the most stable one. The
  mutual information can, however, also be high outside the 1:1
  locking regions, because of higher-order locking, especially when
  $\omega_0/\omega>2$. (B-D) The mutual information as a function of
  $\omega_0/\omega$ for different values of the diffusion constant
  $D$, and for three values of the coupling strength
  $\epsilon/\omega$, as indicated by the dashed lines in panel A:
  $\epsilon/\omega = 0.5$ (B), $\epsilon/\omega=1.5$ (C), and
  $\epsilon/\omega = 4.5$ (D). For all values of $\epsilon$, the mutual
  information increases as $D$ decreases. The peaks outside the main
  locking region around $\omega_0 \approx \omega$ correspond to higher
  order locking.}
\flabel{MutInfo}
\end{figure*}

The results of \fref{MutInfo}A are further elucidated in panels B-D,
which show the mutual information as a function of $\omega_0/\omega$
for different values of the diffusion constant $D$, and for three
different values of $\epsilon/\omega$, respectively; the results for
$D=0.1/T$ in the panels B-D correspond to three different cuts through
the heatmap of panel A. The following points are worthy of
note. First, it can be seen that for each value of $\epsilon/\omega$
and $\omega_0/\omega$ the mutual information always increases with
decreasing $D$. Decreasing the noise makes the mapping from the time
to the phase of the clock more deterministic, which means that the
time can be more accurately inferred from the phase of the clock.  Secondly, 
it is seen that the mutual information exhibits very characteristic
peaks, which result from higher order locking. For example, the peak at $\omega_0/\omega \approx
2.3$ for $\epsilon = 1.5 \omega$, corresponds to 2:1 locking.

\begin{figure*}[t]
\centering
\includegraphics[width=18cm]{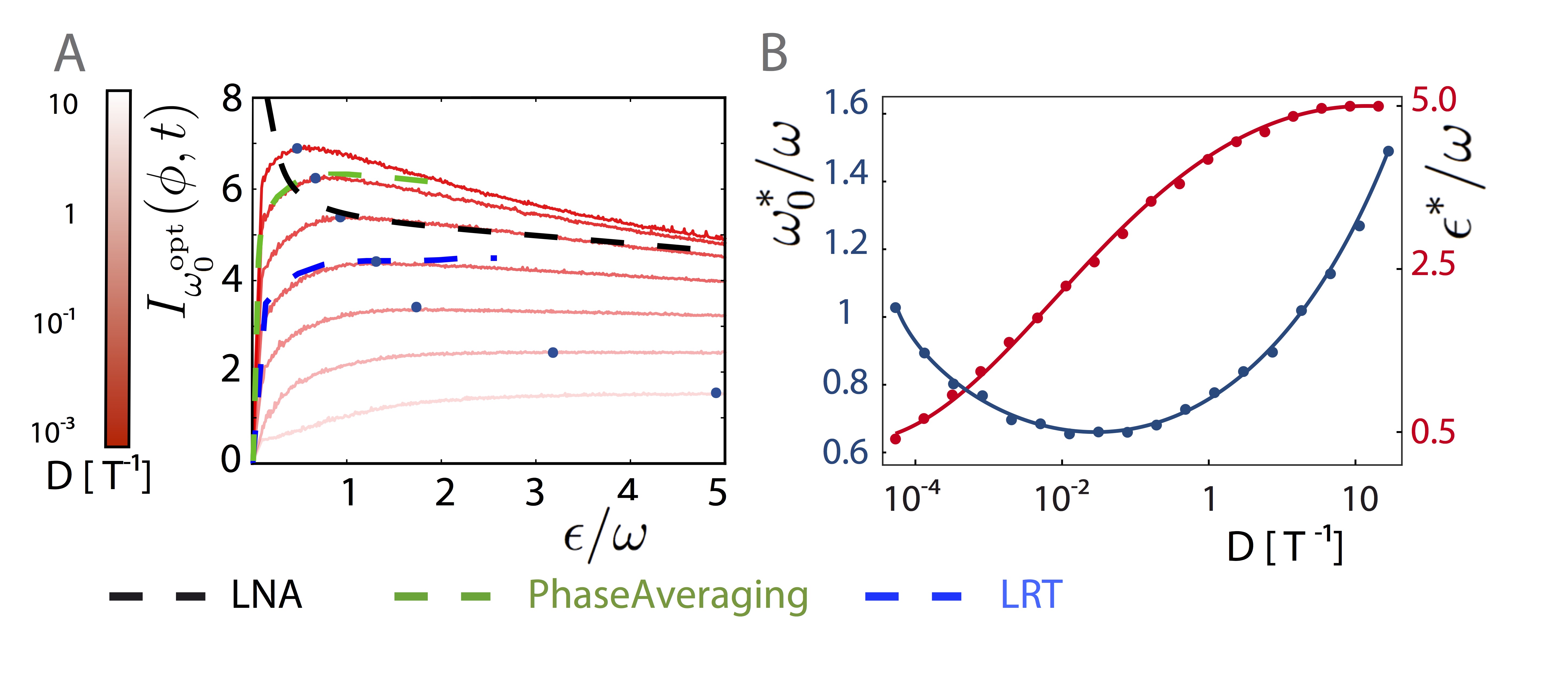}
\caption{Optimal design of the clock: 
  parameters $\epsilon=\epsP=\epsM$ and $\omega_0$ of the
  phase-response curve $Z(\phi)$ that maximize the mutual information
  $I(\phi,t)$ as a function of the intrinsic clock noise $D$, keeping the shape of $Z(\phi)$ constant
  (see \fref{cartoon}A). (A) The mutual information
  $I_{\omega_0^{\rm opt}}(\phi;t)$ obtained by maximizing $I(\phi;t)$
  over $\omega_0$ as a function of $\epsilon$, for different values of
  $D$. It is seen that there is an optimal coupling strength
  $\epsilon_{\rm opt}$ that maximizes the mutual information, which
  depends on the magnitude of the diffusion constant $D$; the blue dot
  denotes the maximum for each value of $D$. The figure
  also shows the predictions of three theories, each for their own
  regime of validity: the linear-noise approximation (LNA), which
  captures the regime of strong coupling $\epsilon$ and low diffusion
  $D$ (result shown for $D=10^{-2}/T$); the phase-averaging method (PAM), which describes the regime of
  weak coupling and weak noise (result shown for $D=10^{-3}/T$); and linear-response theory (LRT),
  which describes the regime of high diffusion and weak coupling
  (result shown for $D=1/T$). For
  a more detailed comparision of the accuries of the respective
  theories, see \fref{KLtest}. (B) The optimal coupling strength
  $\epsilon_{\rm opt}$ (red dots) and the optimal intrinsic frequency
  $\omega_0^{\rm opt}$ (blue dots), both obtained by maximizing $I(\phi;t)$ over {\em both}
  $\epsilon$ and $\omega_0$, as a function of $D$. While
  $\epsilon_{\rm opt}$ increases with $D$ monotonically,
  $\omega_0^{\rm opt}$ first decreases from $\omega_0=\omega$, but then rises
  again to become larger than $\omega$ for higher $D$. The lines are a
  guide to the eye. Other
    parameters: $\Delta \phi_{12}=\Delta \phi_{23}=\pi/2$. }
\flabel{MutInfoOpt}
\end{figure*}

\fref{MutInfo} also shows that, for a given $\omega_0$ and $D$, the mutual
information initially increases with $\epsilon$. This is not
sursurprising, and is consistent with the observation that increasing
the coupling strength $\epsilon$ tends to widen the Arnold Tongue;
locking is enhanced by increasing the coupling strength. However, a
closer examination of the different panels of \fref{MutInfo} suggests
that the mutual information not only saturates as $\epsilon$ is
increased further, but even goes down. The second surprising
observation is that the optimal intrinsic frequency $\omega_0$ that
maximizes the mutual information is not equal to $\omega$. In fact, it
seems to be smaller than $\omega$ when $D$ is small, but then becomes
larger than $\omega$ as $D$ is increased (panel D).

To elucidate the optimal design of the clock that maximizes the mutual
information further, we show in \fref{MutInfoOpt}A the mutual
information $I_{\omega_0^{\rm opt}}(\phi;t)$ that has been obtained by
maximizing $I(\phi;t)$ over $\omega_0$ as a function of $\epsilon$,
for different values of $D$. It is seen that for all values of $D$,
$I_{\omega_0^{\rm opt}}(\phi;t)$ first rises with $\epsilon$, as
expected. However, $I_{\omega_0^{\rm opt}}(\phi;t)$ then reaches a
maximum, after which it comes down: there exists an optimal coupling
strength $\epsilon_{\rm opt}$ that maximizes $I_{\omega_0^{\rm
    opt}}(\phi;t)$; increasing the coupling too much will actually
{\em decrease} the mutual information.  \fref{MutInfoOpt}A also shows,
however, that the optimal coupling $\epsilon_{\rm opt}$ does increase
with the diffusion constant. This is more clearly shown in panel B:
$\epsilon_{\rm opt}$ increases monotonically with $D$.  This panel
also shows the optimal intrinsic frequency $\omega_0^{\rm opt}$
obtained by maximizing the mutual information over both $\omega_0$ and
$\epsilon$, as a function of $D$.  For $D\to 0$, $\epsilon_{\rm opt}$
goes to zero, and $\omega_0$ to $\omega$---this is the free running clock. As
$D$ is increased, however, $\omega_0$ first decreases, but then
increases again to become larger than $\omega$ for higher diffusion
constants. The optimal intrinsic period that maximizes the mutual
information depends in a non-trivial, non-monotonic, manner on the
noise in the clock.

\subsection{Optimal design arises from trade-off between linearity and stability}
To understand the optimal design of the clock, we have to recognize
that, in general, the amount of information that is
transmitted through a communication channel depends on the input
distribution, the input-output relation, and on the noise
that is propagated to the output. For a given amount of noise, the
optimal shape of the input-output relation that maximizes the mutual
information is determined by the shape of the input distribution. 
However, the shape that
optimally matches the input-output curve to the input distribution, is
not necessarily the design that minimizes the noise in the output. Our
system provides a clear demonstration of this general principle, and,
as we will see, the optimal design of the clock can be understood as
arising from a trade-off between stability, i.e. noise minimization,
and linearity, i.e. optimally matching the input-output curve to the
statistics of the input.

When the noise is very weak, noise minimization is not important, and
optimally matching the input-output curve to the input distribution is
paramount. Since the input distribution $p(t)$ is flat, the optimal
input-output curve is linear: the average phase $\overline{\phi}(t)$ should
increase linearly with time $t$.  This is indeed the solution of the free
running clock, $\phi(t) = \omega_0 t$, and it explains why in the
low-noise limit the optimal design is that of an essentially free
running system that is only very weakly coupled to the input.

\begin{figure*}[t]
\centering
\hspace*{-1cm}\includegraphics[scale=0.8]{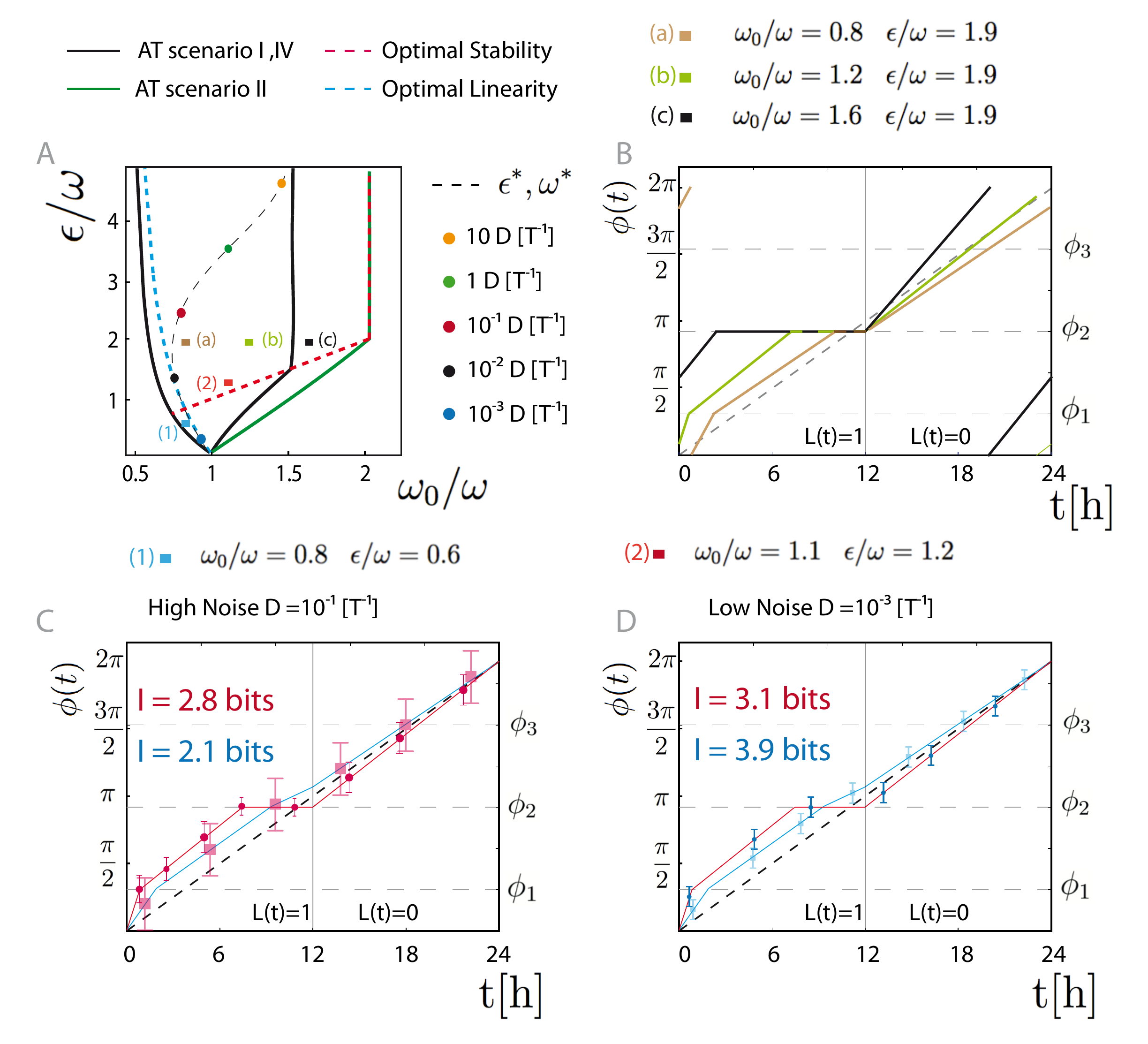}
\caption{The optimal design arises from a trade-off between linearity
  and stability. (A) The black line shows the Arnold Tongue for
    scenario 1 and 4 while the green line shows the Arnold Tongue of
    scenario 2 (see also \fref{ArnoldTong}). The dashed blue line
  shows for each value of $\epsilon$ the value of $\omega_0$ that
  makes the input-output curve, $\overline{\phi}(t)$, most linear,
  i.e. minimizes $\int_0^T dt(\overline{\phi}(t) - \omega t)^2$. The
  dashed red line shows for each value of $\epsilon$ the value of
  $\omega_0$ that maximizes the stability. For $\epsilon/\omega < 2$,
  this line is $\omega_0 = \epsilon$, along which $F^\prime = 0$; for
  $\epsilon = \epsM > \omega_0$, $F^\prime=0$ for all values of
  $\omega_0$ and $\epsilon$; the line of maximal stability then
  corresponds to the line where the system spends most of its time in
  $\phi_2$, which is the line $\omega_0 = \epsilon$ when $\epsilon < 2
  \omega$ and $\omega_0 = 2 \omega$ when $\epsilon \geq \omega$; this
  is further illustrated in panel B. The dashed black line shows a
  parametric plot of the optimal system, i.e. the combination
  $(\epsilon_{\rm opt},\omega^{\rm opt}_0)$ that maximizes the mutual
  information as a function of $D$ (values of $D$ along this solid
  line are indicated by the colored circles; see also \fref{MutInfoOpt}B). It
  is seen that for low diffusion constant, the optimal
  system that maximizes the mutual information (black line) follows
  the dashed blue line where the input-output curve is most linear,
  while for high noise the optimal system moves towards
  the dashed red line, where the system is most stable.  How this
  trade-off between linearity and stability maximizes information
  tranmission is further illustrated in panels (C) and (D). Panel (B)
  shows the average input-output curves for the three points labeled
  (a), (b), and (c) in panel A. It is seen that as the system moves
  towards the line of maximal stability, the time the system spends in
  $\phi_2$ increases; for $\epsilon / \omega > 2$, at $\omega_0 =
  2\omega$, the system starts the day at $\phi_2$. Panel (C) 
  shows the two average input-output curves corresponding to the two
  points (1) and (2) in panel (A), together with the output noise, for a
  high value of the diffusion constant, $D=0.1/T$. Panel (D) shows the
  same, but then for a low value of the diffusion constant,
  $D=10^{-3}/T$. It is seen that when the noise is small (panel D), the
  output noise of the more stable system (red line) is hardly smaller
  than that of the more linear system (blue line); consequently, the
  optimal input-output curve can be linear to maximize information
  transmission. In contrast, when the noise is large (panel C), the
  system with a more linear input-output curve (the blue line) has
  significantly more output noise than the more stable but more
  non-linear system (red line); in this regime, stability becomes
  important for taming the output noise, making the optimal system
  more non-linear (red line).  Other parameters: $\Delta
  \phi_{12}=\Delta \phi_{23}=\pi/2$. \flabel{TradeOff}}
\end{figure*}

However, as the noise level is increased, the reliability by which
each input signal is relayed, becomes increasingly important.  Here, a
trade-off could emerge: while increasing the coupling strength $\epsilon$
could reduce the noise at the output, which tends to enhance information
transmission, it may also distort the input-output curve, pushing it away
from its optimal linear-shape,  decreasing information
transmission. Can we capture this trade-off quantitatively?

To study the trade-off between linearity and stability, we have
computed for each value of $\epsilon$, the value of $\omega_0$ that
makes the average input-output relation $\overline{\phi}(t)$ most
linear, i.e. minimizes $\int_0^T dt (\overline{\phi}(t) - \omega
t)^2$. The result is the blue line in \fref{TradeOff}A, which lies in
the Arnold tongue of scenario I. Along this line of maximal linearity,
$\omega_0$ decreases as $\epsilon$ increases, which can be understood
intuitively by noting that increasing $\epsilon$ introduces a
curvature in the input-output relation, leading to a deviation away
from the straight line $\omega t$: at the beginning of the day, till
the time $t_1$ at which the system crosses $\phi_1$, the phase evolves
with a speed $\omega_0+\epsilon$, whereas between the time $t_2$ at
which the system crosses $\phi_2$ and the end of the day at $T/2$, the
phase evolves either follows $\phi_2$ when $\epsilon=\epsM >
\omega_0$ or evolves with a speed $\omega_0 - \epsilon$ when
$\epsilon=\epsM < \omega_0$. While increasing $\epsilon$ tends to
increase the curvature, this effect can be counteracted by decreasing
$\omega_0$. 

To quantify the stability, we
define the return map $F_t(\phi)$:
\begin{align}
 \phi(t+T)=F_t(\phi(t))=,
\elabel{F}
\end{align}
where the subscript $t$ for $F$ indicates that the return map
  depends on time; this subscript will be suppressed in what follows below when there is no ambguity, in order to simplify notation.
The deterministic solution $\phi^*(t)$ is given by $\phi^*(t) =
\phi^*(t+T) =F(\phi^*(t))$. We now expand $F(\phi)$ around $\phi^*(t)$:
\begin{align}
F(\phi^*+\delta \phi) = F(\phi^*) + F^\prime (\phi^*) \delta
\phi,
\elabel{Fdp}
\end{align}
where $\delta \phi = \phi - \phi^*$ and we have dropped the subscript
$t$ because $F^\prime(\phi)$, which gives the rate of exponential relaxation back to the limit cycle over many cycles, must be independent of $t$.
Indeed, by exploiting that $F(\phi^*(t)) = \phi^*(t+T)$, we find that
\begin{align}
\delta \phi(t+T) = F^\prime(\phi^*) \delta \phi(t).
\elabel{Fprime}
\end{align}
The quantity $F^\prime(\phi^*)\equiv \partial F(\phi)/\partial
\phi|_{\phi^*} = \partial \phi(t+T) / \partial \phi(t)|_{\phi^*(t)}$
determines the linear stability of the system, with $F^\prime < 1$
meaning that the system is stable.  The quantity can be directly
obtained from the deterministic solutions. We first note that, since
$L(t)=0$ during the dark, $F^\prime(\phi^*(t=0))=\partial \phi(T)
/ \partial \phi(0) = \partial \phi(T/2) / \partial \phi(0)$. For
scenario 1, when $\epsM < \omega_0$, $\phi(T/2) =\phi_2 + (\omega_0 -
\epsM) (T/2-t_2)$. We then find that, exploiting \erefstwo{t2}{phis},
$F^\prime(\phi^*(t=0)=\partial \phi(T/2) / \partial \phi(0) = \partial
\phi(T/2) / \partial \phi_s = (\partial \phi(T/2) / \partial t_2)
(\partial t_2 / \partial t_1) (\partial t_1 /\partial \phi_s) =
(\omega_0-\epsM)/(\omega_0 + \epsP)$. Similarly, for scenario 2 we
find that, for $\epsM < \omega_0$,
$F^\prime(\phi^*(t=0)=(\omega_0-\epsM) / \omega_0$. Here, we consider
the case that $\epsM = \epsP = \epsilon$. Clearly, in both scenarios
the stability is maximized when $\epsilon$ approaches $\omega_0$ and
$F^\prime(\phi^*)$ becomes zero. This defines the line
$\epsilon=\omega_0$, along which $F^\prime(\phi^*)=0$; it is the part
of the red dashed line of maximal stability in \fref{TradeOff}A that
corresponds to $\epsilon < 2\omega$.

For $\epsilon = \epsM > \omega_0$, $F^\prime (\phi^*)=0$ for both
scenarios I and II, because during the day the phase evolution of the
system comes to a standstill at $\phi_2$; any perturbation in $\phi$
will fully relax during one period. Can we nonetheless differentiate
in the stability strength, even though the {\em linear} stability
$F^\prime(\phi^*)=0$ for all points $(\epsilon,\omega)$ above the line
$\epsilon = \omega_0$?  To answer this question, we turn to a global
stability measure, which is defined by the amount of time the
deterministic system spends at $\phi_2$, which is the bottom of the
potential well when $\epsilon=\epsM\geq\omega_0$ (see \fref{cartoon}).
The value of $\omega_0$ that maximizes the stability for a given
$\epsilon$ according to this measure, is $\omega_0=\epsilon$ when
$\epsilon \leq 2\omega$ and $\omega_0 = 2 \omega$ when $\epsilon \geq
2 \omega$. This fully specifies the line of maximum stability shown in
\fref{TradeOff}A. The reason why the stability is maximized along this
line, is illustrated in \fref{TradeOff}B. During the night, the
trajectories evolve freely, and because of noise they will arrive at
the beginning of the day with a distribution of phases. Along the line
of maximum stability, the stochastic trajectories are most likely
to reach the bottom of the potential well at $\phi_2$ during the day
(see \fref{cartoon}), where they will be confined, before they are
released again during the night. Indeed, along this stability line the
variance in the phase, $\avg{\delta \phi^2}$, will be lowest which
tends to increase information transmission. However, the input-output
relation $\overline{\phi}(t)$ is then higly non-linear. In fact, the
globally most stable solution, for all possible values of $\epsilon$
and $\omega_0$, is
\begin{align}
 \phi^{\rm stab}(t)\equiv \phi_2 \theta
(T/2-t) + \omega_0 t \theta(t-T/2), \,{\rm with}\,\, \omega_0 = 2 \omega,\elabel{phi_stab}
\end{align}
which is the most stable solution for any $\epsilon \geq 2 \omega$. It
is shown in \fref{TradeOff}B---it is the solution at the
high-frequency boundary of the AT tongue of scenario 2. This solution
maximizes the probability that trajectories that start of the limit
cycle at the beginning of the day, will return to the limit cycle
$\phi_2$ before the end of the day. While this solution is maximally
stable, no time points $t$ can be inferred from $\phi(t)$ during the
day, because $\overline{\phi}(t)$ is completely flat. This
dramatically reduces information transmission.

The optimal values of $\omega_0$ and $\epsilon$ that maximize the
mutual information as a function of the noise in the system can now be
understood as a trade-off between linearity and stability.  This
trade-off is illustrated in the bottom panels of \fref{TradeOff},
which show the average input-output curves, together with their output
noise, for the two points 1 and 2 in the map of panel A, both for a
high diffusion constant (panel C) and a low diffusion constant (panel
D). When the diffusion constant is low (panel D), the noise in the
more stable but more non-linear system (red line, corresponding to
point 2) is hardly lower than that in the more linear but less stable
system (blue line, corresponding to point 1), which means that the
benefit of linearity dominates and the mutual information is maximized
in the more linear system. In contrast, when the noise is larger
(panel C), the output noise in the more stable but more non-linear
system (red line) is so much smaller than that in the less stable but
more linear system (blue line) that it outweighs the cost of higher
non-linearity, thus maximizing mutual information. 

Finally, panel A also shows a parametric plot of the optimal
$(\epsilon,\omega_0)$ that maximizes the mutual information, with the
noise $D$ the parameter that is being varied (dashed black line; the
colors of circles denote values of the diffusion
constant). It is seen that for low $D$ the optimal system traces the
dashed blue line of maximal linearity, but then at a higher $D$ makes a
transition towards the dashed red line line of maximal stability.

\subsection{The optimal shape of the phase response curve}
In the previous section, we showed how the optimal values of the
coupling strength $\epsilon$ and the intrinsic frequency $\omega_0$
depend on the noise $D$ in the system, while keeping the shape of the
coupling function $Z(\phi)$ constant.  In this section,
we will relax this restriction.

We first checked the effect of changing the magnitude of the positive
and negative lobe of the coupling function $Z(\phi)$ as characterized
by $\epsP$ and $\epsM$, respectively (see \fref{cartoon}), keeping
$\Delta \phi_{12}=\Delta \phi_{23}=\pi/2$ constant. We varied
$\epsP$ and $\epsM$ via a parameter $\alpha$, defined as $\epsP =
(1-\alpha) \epsilon$ and $\epsM = \alpha \epsilon$; changing $\alpha$
thus keeps the total absolute coupling strength (the integrated
modulus) constant. We found, however, that the results are not very sensitive to
the precise values of $\epsP$ and $\epsM$ (see Appendix \ref{sec:EpsPepsM}).

\begin{figure}[t]
\centering
\includegraphics[width=9cm]{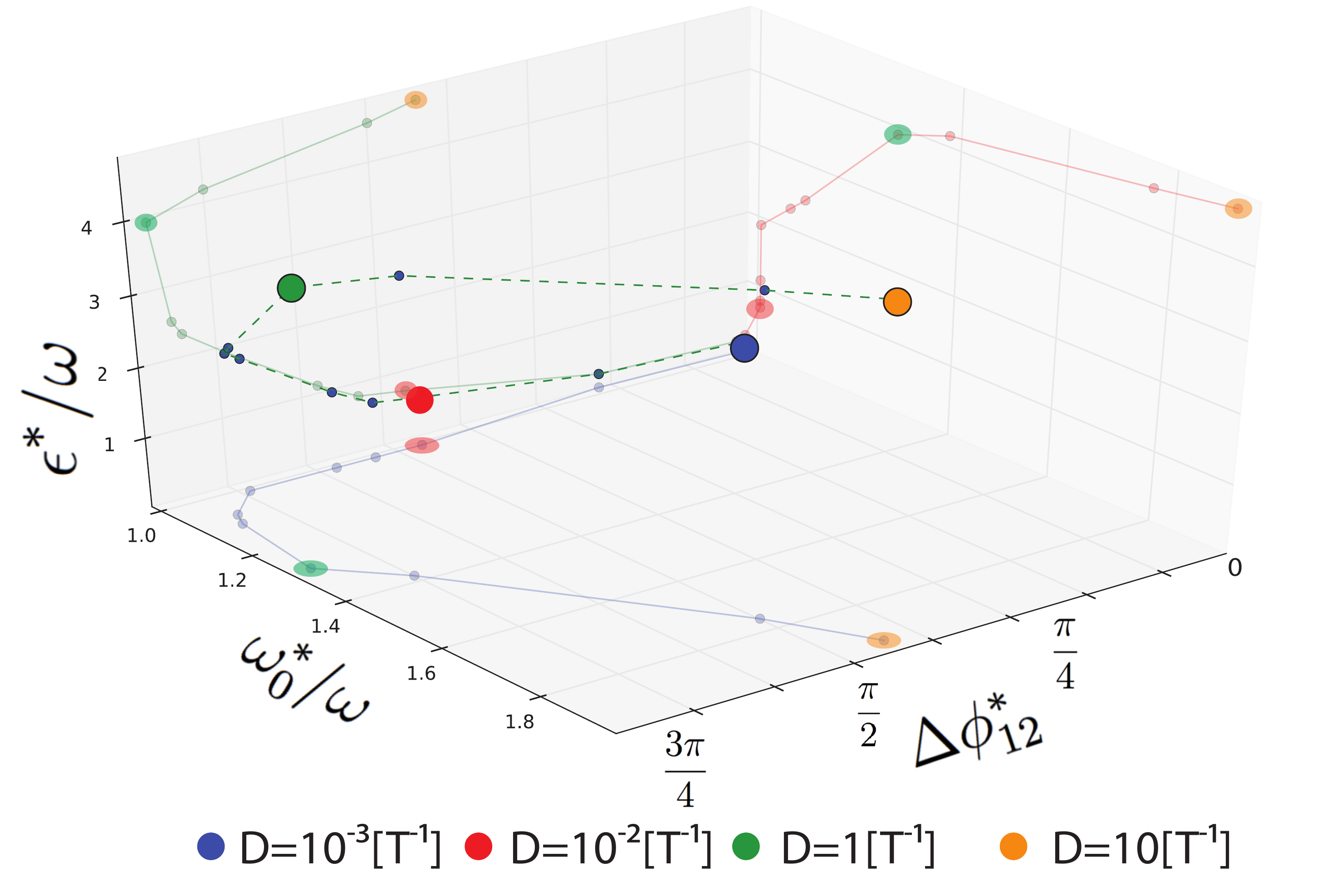}
\caption{A paramteric plot of the optimal coupling strength
  $\epsilon^*(D)$, the optimal intrinsic frequency $\omega_0^*(D)$ and
  the optimal width of the deadzone $\Delta \phi_{12}^*(D)$ that
  maximize the mutual information, with the noise $D$ being the
  parameter that is varied. The value of $\Delta \phi_{23}=\pi/2$ was
  kept constant. It is seen that $\epsilon^*$ rises with $D$, while
  $\omega_0^*$ remains initially close to $\omega$, but then rises
  too. In contrast, $\Delta \phi_{12}^*$ first increases and then
  decreases. Colored dots give the diffusion constants for which
    $(\epsilon^*,\omega_0^*,\Delta \phi_{12}^*)$ are
    optimal.
  \flabel{GlobOpt}}
\end{figure}

We then decided to compute the mutual information $I(\phi,t)$ as a
function of $\Delta \phi_{12}$ and $\Delta \phi_{23}$ for different
values of $\epsilon$, $\omega_0$, and $D$, keeping $\epsP=\epsM =
\epsilon$. We found that the mutual information is essentially
independent of $\Delta \phi_{23}$. This can be understood as follows: The
deterministic Arnold tongue and, to a good approximation, the dynamics
of the stochastic system, does not depend on the absolute values of
$\phi_1,\phi_2,\phi_3$, but only on $\Delta \phi_{12}$ and $\Delta
\phi_{23}$ (see section \ref{sec:AT}).  Moreover, as long as $\phi_3$
is crossed during the night (see \fref{cartoon}), we can change
$\phi_3$ at will, because during the night, when $L(t)=0$, the clock
is not coupled to light (see \eref{dphidt}), meaning that the clock
runs with its intrinsic frequency $\omega_0$. Changing $\Delta
\phi_{23}$ by changing $\phi_3$ will thus have no effect. Changing
$\Delta \phi_{23}$ by changing $\phi_2$ will also have no effect when
$\phi_1$ is simultaneously changed such that $\Delta \phi_{12}$ remains
constant: while changing $\phi_2$ and $\phi_1$ keeping $\Delta
\phi_{12}$ and $\phi_3$ constant will alter $\Delta \phi_{23}$, we can
always change $\phi_3$ such that $\Delta \phi_{23}$ remains
unchanged. In short, as long as $\phi_3$ is crossed during the night
(which it will for most values of $\phi_1$ and $\phi_2$), changing
$\phi_1$ and $\phi_2$ keeping $\Delta \phi_{12}$ constant, does not
change the dynamics; the times $t_1$ and $t_2$ at which $\phi_1$ and
$\phi_2$ are crossed, respectively, do not change.

\begin{figure*}[t]
\centering
s\includegraphics[width=16cm]{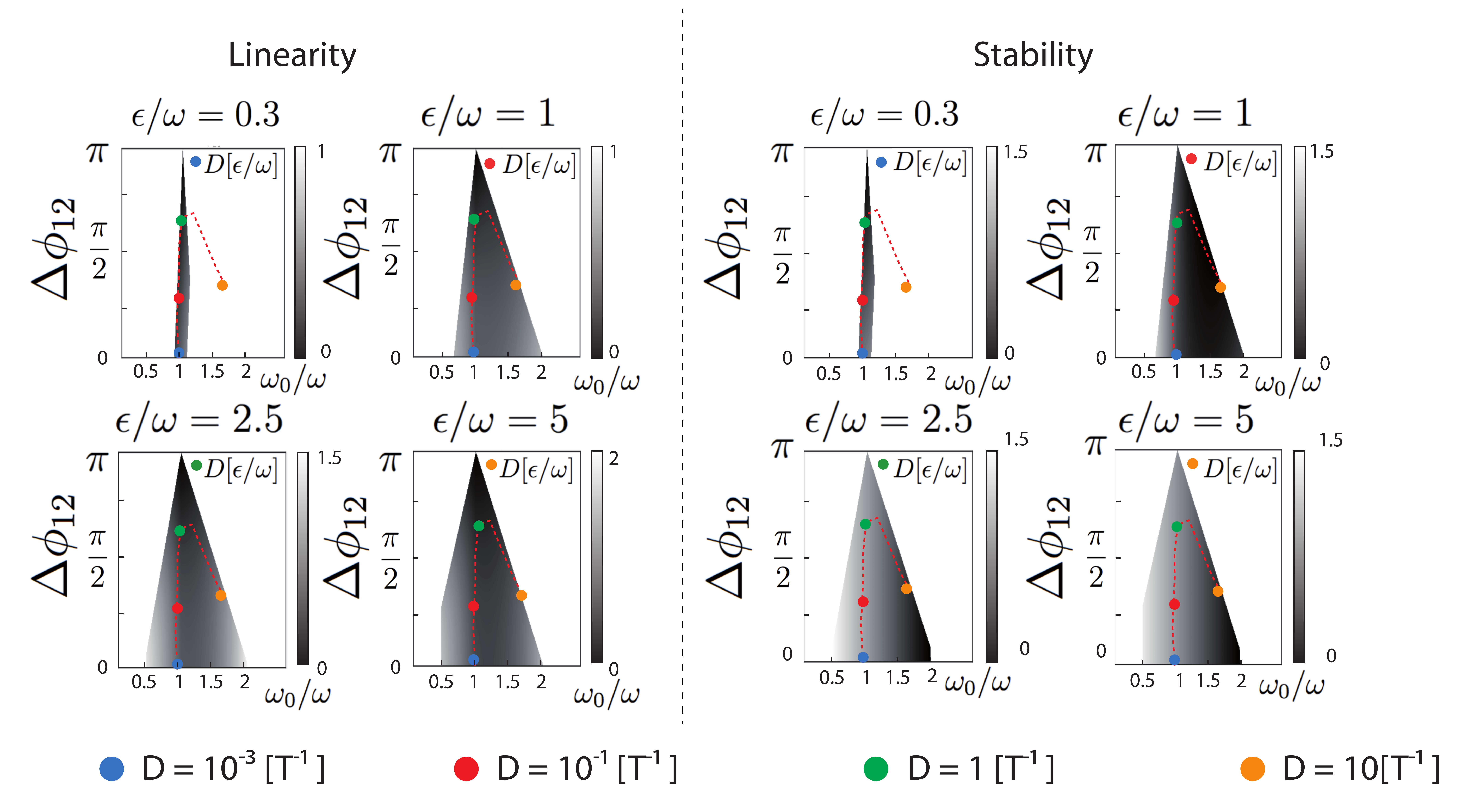}
\caption{The optimal shape of the instantaneous phase response curve
  $Z(\phi)$ arises as a trade-off between linearity and stability.
  The linearity (A) is quantified via $\int_0^T dt (\overline{\phi}(t)
  - \phi^{\rm lin}(t))^2$, which is the average deviation of the mean
  input-output relation $\overline{\phi}(t)$ away from the most linear
  solution $\phi^{\rm lin}(t)=\omega t$. The stability (B) is
  quantified via $\int_0^T dt(\overline{\phi}(t) - \phi^{\rm
      stab}(t))^2$, which is the average deviation of
    $\overline{\phi}(t)$ away from the most stable solution $\phi^{\rm
      stab}(t)$, given by \eref{phi_stab}. These measures are
  computed as a function of the intrinsic frequency $\omega_0$ and the
  width of the deadzone $\Delta \phi_{12}$, for different values of
  $\epsilon$, inside the Arnold Tongue of scenario 1; note that
  smaller values correspond to higher linearity and stability,
  respectively.  Superimposed is a parametric plot of the optimal
  intrinsic frequency $\omega_0^*(D)$ and optimal width of the
  dead-zone $\Delta \phi_{12}^*(D)$ that maximize the mutual
  information for a given $D$. The dots denote the values of $D$ to
  which $\omega_0^*(D)$ and $\Delta \phi_{12}^*(D)$ correspond; the value
  of $D$ for which the $\epsilon$ of a panel is the optimal coupling
  strength $\epsilon^*$ is given near the top of the Arnold tongue. It
  is seen that for small $D$, the optimal parameters
  $(\omega_0^*(D),\Delta \phi_{12}^*(D),\epsilon^*(D))$ that maximize
  the mutual information are those that make the input-output relation
  $\overline{\phi}(t)$ most linear (top left panel A), while for large
  $D$, the optimal parameters are those that make the system very
  stable (bottom right panel B).  Other parameters: $\Delta
  \phi_{23}=\pi/2$. \flabel{DeltaPhi12}}
\end{figure*}

Because $\phi_{23}$ is not critical, we kept $\Delta \phi_{23}=\pi/2$,
and then performed very extensive simulations to determine the optimal
coupling strength $\epsilon^*$, speed $\omega_0^*$ and optimal
deadzone $\Delta \phi_{12}^*$ that maximize the mutual information, as
a function of $D$. \fref{GlobOpt} shows a parametric plot of
$\epsilon^*(D), \omega_0^*(D)$ and $\Delta \phi_{12}^*(D)$, with $D$
being the parameter that is varied.
 It is seen that for very low $D$, the optimal coupling
strength $\epsilon^*$ is small, the optimal intrinsic frequency
$\omega_0^*$ is close to $\omega$, and the optimal value of $\Delta
\phi_{12}^*$ is small. As the diffusion constant is increased, $\epsilon^*$
rises but $\omega_0^*$ initially remains close to $\omega$ and then
increases too.  The optimal value of $\Delta \phi_{12}$, however, first
rises and then falls again.

The behavior of $\Delta \phi_{12}^*$ can again be understood as a
trade-off between linearity and stability. This is illustrated in
\fref{DeltaPhi12}. The figure shows for different values of $\epsilon$
the linearity and the stability of the input-output relation
$\overline{\phi}(t)$ as a function of $\Delta \phi_{12}$ and
$\omega_0$, computed within the deterministic Arnold tongue of
scenario 1 (where the mutual information is highest). The linearity of
$\overline{\phi}(t)$ is quantified via $\int_0^T dt
(\overline{\phi}(t) - \phi^{\rm lin}(t))^2$, which is the average
deviation of $\overline{\phi}(t)$ away from the most linear
input-output relation, $\phi^{\rm lin}(t) = \omega t$. The stability
of $\overline{\phi}(t)$ is quantified via $\int_0^T dt
(\overline{\phi}(t) - \phi^{\rm stab}(t))^2$, which is the average
deviation of $\overline{\phi}(t)$ away from the most stable
input-output relation $\phi^{\rm stab}(t)$, given by \eref{phi_stab}.

 The following observations can be made. First, the width of the
Arnold tongue (the range of $\omega_0$ that permits a deterministic
solution) decreases as $\Delta \phi_{12}$ increases. Secondly, the
linearity is maximal in the range $1 < \omega_0/\omega < 1.5$, and
tends to increase with $\Delta \phi_{12}$: in the deadzone
$\Delta \phi_{12}$ the system evolves freely with speed
$\omega_0$, which makes $\overline{\phi}(t)$ more linear, especially when $\omega_0 \sim
\omega$. In contrast, the stability is highest when $\omega_0/\omega$
is large and $\Delta \phi_{12}$ is small, particularly for higher values
of $\epsilon$. The large magnitudes of $\omega_0$ and $\epsilon$ mean
that at the beginning of the day the system is strongly driven,
$\avg{\dot{\phi}}\approx \epsilon+\omega_0$, and the small deadzone
$\Delta \phi_{12}$ means that after the system has crossed $\phi_1$,
it quickly reaches $\phi_2$, where, with $\epsilon = \epsM >\omega_0$, the system is then confined (see
\fref{cartoon}). 

\fref{DeltaPhi12} also shows superimposed a parametric plot of the optimal
$\Delta \phi_{12}^*(D)$ against the optimal $\omega_0^*(D)$.  The
colored dots denote the diffusion constants for which
$(\omega_0^*,\Delta \phi_{12}^*)$ are optimal; the diffusion constant
for which the $\epsilon$ of a panel is the optimal coupling strength
$\epsilon^*$ is shown near the top of the Arnold tongue. It is seen
that for very small $D$, the optimal system parameters
$(\omega_0^*,\Delta \phi_{12}^*, \epsilon^*)$ put the system in the
regime where $\phi(t)$ is linear (top left panel A); increasing
$\Delta \phi_{12}^*$ would not make the system significantly more
linear, since $\epsilon^*$ is still very small. Increasing $D$ raises
$\epsilon^*$, while $\omega^*$ remains close to $\omega$. The optimal
width of the deadzone $\Delta \phi_{12}^*$ now increases, because for
the higher value of $\epsilon^*$ the system becomes significantly more
linear when $\Delta \phi_{12}^*$ is increased. Beyond $D=1/T$,
however, linearity is sacrificed for stability. The optimal coupling
strength $\epsilon^*$ and intrinsic frequency $\omega^*$ increase,
while the optimal size of the deadzone decreases, to maximize
stability. Indeed, when the noise is even larger still, the width of
the deadzone reduces to zero and the coupling strength and intrinsic
frequency become even larger: during the day the system is rapidly
driven to $\phi_2$, where it then remains strongly confined till the
beginning of the night (see
\fref{cartoon} and also \fref{TradeOff}C). In this limit, the clock
transmits one bit of information, and the system can only
distinguish between day and night.

\fref{GlobOpt} thus generalises the finding of \fref{TradeOff} that
corresponds to a fixed deadzone and shows that the optimal shape of
the instantaneous phase response curve can be understood as a
trade-off between linearity and stability.

\section{Theory}
The simulation results can be described quantitatively via three
different theories, which each accurately describe a particular regime
of parameters:  The
linear-noise approximation (LNA) describes the regime of strong
coupling and low diffusion; the phase-averaging method (PAM) holds in the low
diffusion, weak coupling regime; and the linear-response theory (LRT) applies in the regime of high noise and weak coupling. Here, we have borrowed the terminology
  LNA from the name of the theory to describe biochemical networks
  that is based on the same underlying principles: indeed, rather than
  linearizing the Chemical Langevin Equation around the fixed point
  given by the mean-field chemical rate equations and taking the noise
  at that fixed point, we here linearize the return map $F(\phi)$
  around its fixed point, and compute the noise at that fixed point.
The results of the respective theories in their regime of validity are
shown in \fref{MutInfoOpt}. A more detailed comparison between the
simulation results and the theoretical predictions, discussed below,
is shown in \fref{KLtest}, where $\epsilon$ and $D$ are varied for two
different values of $\omega_0$.

\subsection{Linear-Noise Approximation}
The linear-noise approximation (LNA) is expected to be accurate when
the driving is strong compared to the diffusion constant, so that the
system closely follows the deterministic solution $\phi^*(t)$, which
is given by the return map of \eref{F}: $\phi^*(t) = \phi^*(t+T)
=F(\phi^*(t))$. Because in this regime the deviations from the
deterministic solution are small, we can expand $F(\phi)$ up to
linear order in $\delta \phi = \phi - \phi^*$ to obtain
$F(\phi^*+\delta \phi)$, see \eref{Fdp}. This makes it possible
to derive how a deviation from the deterministic solution at time $t$
will relax to the limit cycle at time $t+T$: $\delta \phi(t+T) = F^\prime (\phi^*) \delta \phi(t)$
(see \eref{Fprime}). The quantity $F^\prime(\phi^*)$ thus determines
the stability of the system near the deterministic fixed point. It
can be readily obtained from the deterministic solutions.  

Given a variance at time $t$, $\avg{\delta \phi(t)^2}$, the variance
 at time $t+T$,  $\avg{\delta \phi(t+T)^2}$, is given by two
 contributions:
\begin{align}
\avg{\delta \phi(t+T)^2} = F^{\prime^2}(\phi^*) \avg{\delta \phi(t)^2}
+ V\left[\phi(t+T)|\phi^*(t))\right].
\elabel{delphisqLNA0}
\end{align}
The first contribution is a deterministic contritbution, which is
determined by how a deviation $\delta \phi(t) = \phi(t) - \phi^*(t)$
at time $t$ regresses deterministically to the mean at time $t+T$:
$\delta \phi(t+T)= F^\prime (\phi^*) \delta \phi(t)$. The second
  contribution describes the variance of the distribution
  $P(\phi(t+T)|\phi^*(t))$ of $\phi(t+T)$ at time $t+T$, given
  that at time $t$ the system was at the deterministic solution
  $\phi^*(t)$; in general, we should instead compute the variance at $t + T$ for an arbitrary initial $\phi(t) = \delta \phi(t) + \phi^*(t)$, but to leading order in small $\delta \phi$ it is sufficient to evaluate the noise at the deterministic solution $\phi^*$.
 It is important to
  note that the variance $V\left[\phi(t+T)|\phi^*(t))\right]$ depends
  not only on the diffusion constant, but also on the deterministic
  force, as in a canonical LNA description: For example, in the simplest possible noisy dynamics, $\dot{\delta x} = -k
  \delta x (t) + \eta (t)$, with $\avg{\eta (t) \eta(t^\prime)} = 2D
  \delta (t-t^\prime)$, the deterministic contribution to the variance
  $\avg{\delta x(t+T)^2}$ at time $t+T$, given the variance
  $\avg{\delta x(t)^2}$ at time $t$, is $\avg{\delta x(t)^2}e^{-2
    kT}$, while the stochastic contribution to the variance at time
  $t+T$ is $V[\delta x(t+T)|x^*(t)]=(D/k) (1-e^{-2 kT})$, which indeed
  depends on the force constant $k$. However, in the limit that the
  force is weak, the stochastic contribution is given by the variance
  of free diffusion: $V[\delta x(t+T)|x^*(t)]= 2D
  T$.  We assume, and subsequently verify numerically, that a similar simplification applies for our phase oscillator model.  Indeed, except at the boundaries $\phi_1$, $\phi_2$, and $\phi_3$, our phase dynamics reduces to diffusion with a constant drift, for which it is rigorously true that $V\left[\phi(t+T)|\phi^*(t))\right] = 2 D T$; our assumption hence amounts to neglecting any corrections to the integrated noise due to the brief ``kicks'' at these boundaries.
  \eref{delphisqLNA0} then reduces to
\begin{align}
\avg{\delta \phi(t+T)^2} = F^{\prime^2}(\phi^*) \avg{\delta \phi(t)^2}
+ 2 D T.
\elabel{delphisqLNA}
\end{align}
This expression constitutes the fluctuation-dissipation relation for
this system. 
In steady state, $\avg{\delta \phi(t+T)^2}=\avg{\delta \phi(t)^2}$,
from which it follows that
\begin{align}
\avg{\delta \phi(t)^2} = \frac{2D T }{1-F^{\prime^2} (\phi^*)}.
\elabel{varLNA}
\end{align}
Clearly, the variance depends not only on the diffusion constant, but
also on the stability, which increases with
the coupling strength; as derived below \eref{Fprime}, for scenario 1,
$F^\prime(\phi^*) = (\omega_0 - \epsM) / (\omega_0 + \epsP)$ decreases
(meaning the system becomes more stable) as $\epsM$ and $\epsP$ increase.

In this linear-noise approximation, the distribution of the phase at
time $t$ is a simple Gaussian with a mean $\overline{\phi}(t)$ that is given by the
deterministic solution, $\overline{\phi}(t) = \phi^*(t)$, and a variance
  that is given by \eref{varLNA}:
\begin{equation}
P(\phi|t) = \frac{1}{\sqrt{2 \pi \sigma_{\phi}}} \exp-\frac{(\phi-\overline{\phi}(t))^2}{2\sigma_{\phi}^2},
\end{equation}
where $\sigma_\phi \equiv \sqrt{\avg{\delta \phi^2}}$. This variance
is, in this approximation, independent of the phase. 

To derive the mutual information, it is conventient to invert the
problem and look for the distribution of possible times $t$, given
$\phi$. This can be obtained from Bayes' rule:
\begin{equation}\label{Ba_ru}
P( t |\phi) = P(t) \frac{P(\phi|t)}{P(\phi)}
\end{equation}
where $P(t)=1/T$ is the uniform prior probability of having a certain
time and $P(\phi)$ is the steady state distribution of $\phi$,
which in the small noise limit can be computed via $P(t) dt =
P(\phi) d\phi$. 
If the noise $\xi$ is small compared to the mean, then $P(t|\phi)$ will
be a Gaussian distribution that is peaked around $t^*(\phi)$, which is
the best estimate of the time given the phase
\cite{Tkacik2011,Dubuis2013,Monti2016}:
\begin{equation}
P(t|\phi)\simeq  \frac{1}{\sqrt{2 \pi \sigma_t^2}} \exp\left[
-\frac{(t-t^*(\phi))^2}{2 \sigma_t^2}\right].
\end{equation}
Here $\sigma^2_t=\sigma^2_t(t^*)$ is the variance in the estimate of the
time, and it is given by \cite{Dubuis2013}
\begin{equation}\elabel{sigt}
\sigma_t^2 = \sigma_{\phi}^2 \left(\frac{dt}{d \overline{\phi}}\right)^2.
\end{equation}
We note that $\sigma^2_t$ does depend on $t$ because the slope
$d\overline{\phi}/dt$ depends on $t$. Indeed, while the LNA assumes
that $\sigma^2_\phi$ is independent of $\phi$, it does capture the
fact that changing $\epsilon$ and $\omega_0$ can affect the mutual
information not only by changing the noise $\sigma^2_\phi$ but also
via the slope $d\overline{\phi}/dt$ of the input-output relation
$\overline{\phi}(t)$.

The mutual information can now be obtained from:
\begin{align}
I(\phi;t) &= H(t) - \left \langle H(t|\phi) \right \rangle_\phi\\
&=\log_2 T -  \left\langle \frac{1}{2}\log_2\left( 2\pi e \sigma_\phi^2
  \left(\frac{dt }{d\overline{\phi}}\right)^2\right) \right\rangle_{\phi}\\
&=\log\left(\frac{T}{\sqrt{2\pi e \sigma^2_\phi}}\right) + \frac{1}{T}\int_0^T dt \log \frac{d\overline{\phi}}{dt},
\label{LNAInfo}
\end{align}
where $\langle \dots \rangle_\phi$ denotes an average over $P(\phi)$,
and we have exploited that in the LNA the variance $\sigma^2_\phi$ is
independent of $\phi$. For the model presented here,
$\overline{\phi}(t)=\phi^*(t)$ is piecewise linear, and the second
integral can be obtained analytically, for each of the scenarios; for
scenario 1, for example, the second term is $1 / T \left(t_1 \log
  (\omega_0 + \epsP) + (t_2 - t_1) \log \omega_0 +\right.$\linebreak
$\left.(T/2-t_2) \log (\omega_0 - \epsM) + T/2 \log \omega_0\right)$.

\fref{MutInfoOpt} shows that the LNA accurately predicts the mutual
information $I_{\omega_0^{\rm opt}}(\phi;t)$ in the regime that the
coupling strength $\epsilon$ is large and the diffusion constant $D$
is small. A more detailed comparison is shown in \fref{KLtest}, which
shows the Kullback-Leibler divergence $D_{KL}(P_n||P_a)$ between the
distribution $P_n = P_n(\phi|t)$ obtained in the simulations and $P_a
= P_a(\phi|t)$ as predicted by LNA. Panels A and B show the result for
$\omega_0/\omega = 1$, while panels C and D show the results for
$\omega_0/\omega  =1.05$. Moreover, panels A and C show the results as
a function of $D$ for two values of $\epsilon$, while panels B and D
show the results as a function of $\epsilon$ for two values of $D$.

Panels A and C show that as $D$ is decreased at fixed $\epsilon$,
  the LNA becomes accurate for small $D$, as expected. Panels B and D
  show that for large $D$, the LNA never becomes accurate, even for
  large $\epsilon$. However, for large values of $\epsilon$, the
  assumption that the stochastic contribution to the variance is given
  by that of free diffusion, $V[\delta \phi(t+T) | \phi^*(t)] \simeq 2
  DT$, breaks down. This is also the reason why for the smaller value
  of $D$ (crosses in panels B and D), the LNA works very well for low
  values of $\epsilon$, but then becomes slighly less accurate for
  higher values of $\epsilon$. Indeed, for $\epsilon=\epsM >
  \omega_0$, $F^\prime = 0$, and the key assumption of
    LNA---namely that the dynamics can be expanded to linear order
    around the deterministic fixed point---breaks down.

  Comparing panel C against panel A and panel D against panel B shows
  that LNA is less accurate in the small $D/\epsilon$ regime when
  $\omega_0/\omega=1.05$ (panels C/D) than when $\omega_0/\omega =
  1.0$ (panels A/B). More specifically, while LNA is very accurate for
  $D < 10^{-2}/T$ for both values of $\epsilon$ when $\omega_0/\omega
  = 1.0$ (panel A), LNA becomes less accurate for $D<10^{-2}/T$ when
  $\omega_0/\omega = 1.05$ and $\epsilon$ is small, {\it i.e.}
  $\epsilon/\omega=0.1$ (panel C); only for $\epsilon/\omega = 0.9$ is
  LNA still accurate in this regime. Similarly, while LNA is very
  accurate for $\epsilon/\omega < 1$ when $D=10^{-3}/T$ and $\omega_0/\omega
  = 1.0$ (panel B), LNA becomes less accurate for $\epsilon/\omega
  < 0.5$ when $D=10^{-3} / T$ yet $\omega_0/\omega = 1.05$ (panel
  D). This observation can be understood by noting that when
  $\omega_0$ is increased, the system moves to the boundary of the
  Arnold Tongue of scenario I, especially when $\epsilon$ is small
  (see \fref{ArnoldTong}). The system then switches under the
  influence of noise between the solution of scenario I and that of
  scenario II, meaning that the response becomes non-linear and LNA
  breaks down. Interestingly, however, another method, described in
  the next section, accurately describes this regime.  
\begin{figure*}[t]
\centering
\includegraphics[scale=0.65]{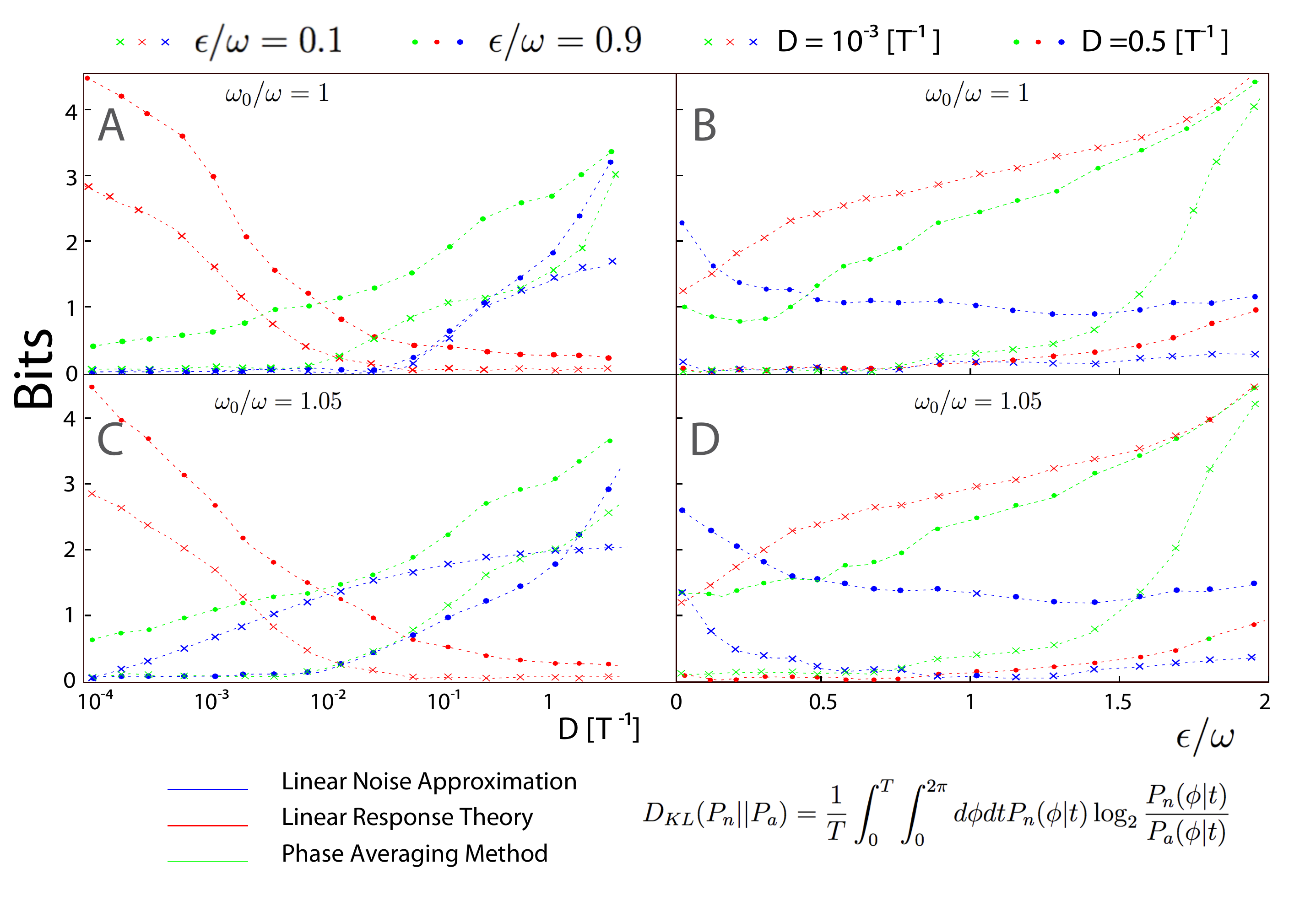}
\caption{Comparison between simulation results and three different
  theories: linear-noise approximation (LNA), phase-average method
  (PAM), and linear-response theory (LRT). The comparison is performed
  by computing the Kullback-Leibler divergence $D_{KL}(P_n||P_a)$ between $P_n(\phi|t)$
  as obtained in the simulations and $P_a(\phi|t)$ as predicted by the
  theory. For two values of $\omega_0$, namely
    $\omega_0/\omega=1$ (panels A,B) and $\omega_0/\omega=1.05$
    (panels C,D), we show $D_{KL}(P_n||P_a)$ as a function of $D$ for two
    values of $\epsilon$ (panels A,C) and $D_{KL}(P_n||P_a)$ as a function
    of $\epsilon$ for two values of $D$ (panels B,D). It is seen that the LNA accurately predicts the regime of
  strong coupling and low noise; PAM the regime of weak coupling and
  weak noise; and LRT the regime of high noise and weak coupling.
 Other parameters: $\Delta
  \phi_{12}=\Delta \phi_{23}=\pi/2$ for all data points.
  \flabel{KLtest}}
\end{figure*}

\subsection{Phase-Averaging Method}
  In the limit that the coupling $\epsilon$ is weak, the
diffusion constant $D$ is small, and the intrinsic frequency
$\omega_0$ is close to the driving frequency $\omega$, we expect that
the evolution of $\phi$ is close to that of the free-running
oscillator, $\phi_0(t) = \omega_0 t + \phi_0$. In this regime the
phase will exhibit fluctuations that are slow, occurring on time
scales much larger than the intrinsic period $T_0$. The detailed
coupling within a clock cycle becomes irrelevant, and only the average
coupling over a clock period matters. This leads to the notion of
phase averaging, in which $P(\phi(t)-\omega t|t)$ no longer depends on
$t$: $P(\phi(t)-\omega t|t) = P(\phi(t)-\omega t) \equiv P(\psi)$,
with $\psi \equiv \phi(t) - \omega t$.

Following Pikovsky \cite{Pikovsky2003}, we now make this intuitive notion
concrete by rewriting the coupling term as
\begin{align}
Q(\phi,t) &= Z(\phi) L(t) \\
&=\sum_k \sum_l a_k b_l e^{i (k \phi + l \omega t)}. \label{eq:Q_phi_t}
\end{align} 
If the coupling and the noise are weak, $\epsilon \to 0, D \to 0$, we
may expect that $\phi \simeq \omega_0 t + \phi_0$ for all times $t$. If we
substitute this into Eq. \ref{eq:Q_phi_t}, we find
\begin{align}
Q(\phi,t) &= \sum_k \sum_l a_k b_l e^{i k \phi_0}e^{i(k \omega_0 + l
  \omega t)}.\elabel{Qkl}
\end{align}
When $\omega \approx \omega_0$, the terms $k=-l$ contribute most
strongly to the integral. These terms correspond to 
variations in the force on long time scales. We thus expect that in the regime
that $\epsilon, D \to 0$ and $\omega \approx \omega_0$, where the phase is expected to follow $\phi \approx
\omega_0 t + \phi_0$, the terms
$k=-l$ yield the strongest contributions to the force:
\begin{align}
Q(\phi,t) &=\sum_k a_k b_{-k} e^{i k(\phi - \omega t)}\\
&=\int_0^T dt^\prime Z(\psi+\omega t^\prime) L(t^\prime) \elabel{Q_phi_t_psi}\\
&=Q(\psi).
\end{align}
where in Eq. \eref{Q_phi_t_psi} we have introduced the new phase variable $\psi
\equiv \phi - \omega t$. The force $Q(\psi)$ is commonly referred to
as the phase-response curve; it is thus a convolution of the
instantaneous phase-response curve $Z(\phi)$ and the light-signal
$L(t)$. 

The temporal evolution of $\psi$, $\dot{\psi} = \dot{\phi}-\omega$,
is, using
 \eref{dphidt}:
\begin{align}
\frac{d \psi}{dt} &= \omega_0 - \omega + \epsilon Q(\psi) + \xi(t)\\
&=-\nu + \epsilon Q(\psi) + \xi(t),
\label{eq:dpsidt}
\end{align}
with $\nu=\omega - \omega_0$. The first two terms on the right-hand
side are the deterministic force, which can be written as the
derivative of a potential $V(\psi)$
\begin{align}
-\nu + \epsilon Q(\psi) = -\frac{dV(\psi)}{d\psi},
\end{align}
with the potential given by
\begin{align}
V(\psi) = \nu \psi - \epsilon \int_{-\pi}^\psi Q(x) dx.
\end{align}
Indeed, the evolution of $\psi$ can be described as that of a particle in a
potential $V(\psi)$, which is a $2\pi$-periodic potential with a slope given
by $\nu = \omega - \omega_0$.

The evolution of the probability density $P(\psi,t)$ is given by the
Fokker-Planck equation corresponding to Eq. \ref{eq:dpsidt}:
\begin{align}\label{FPeq}
\partial_t P(\psi,t) &= - \partial_\psi \left[(-\nu + \epsilon Q(\psi))
P(\psi,t)\right] + D \partial^2_\psi P(\psi,t)\\
&= - \frac{\partial J(\psi,t)}{\partial \psi},
\end{align}
where we have defined the probability current
\begin{align}
J(\psi,t) = -P(\psi,t) \frac{dV(\psi)}{dx} -D \frac{\partial
  P(\psi,t)}{\partial t}.
\end{align} 

In steady state, $\partial P(\psi,t) / \partial t=0$, which yields the
following stationary
solution that is $2\pi$-periodic in $\psi$:
\begin{align}
\overline{P}(\psi) = \frac{1}{C} \int_\psi^{\psi+2\pi}
e^{\frac{V(\psi^\prime) - V(\psi)}{D}}d\psi^\prime.
\end{align}
Here, $C$ is the normalization constant.

\fref{MutInfoOpt} shows that the phase-averaging method (PAM)
accurately predicts the mutual information $I(\phi;t)$ in the regime that both the coupling strength
$\epsilon$ and the diffusion constant $D$ are small. The more detailed
comparison based on the Kullback-Leibler divergence $D_{KL}(P_n||P_a)$
between the distribution $P_n = P_n(\phi|t)$ obtained in the
simulations and $P_a = P_a(\phi|t)$ as predicted by PAM confirms this
interpretation: as shown in panel B of \fref{KLtest}, when
  $\omega_0/\omega = 1.05$, PAM is accurate for $D<10^{-2}/T$ when
  $\epsilon/\omega=0.1$ (green crosses), while LNA breaks down in this
  regime (blue crosses). Similarly, as illustrated in panel D, when
  $\omega_0/\omega=1.05$, PAM is accurate for $\epsilon/\omega < 0.7$
  when $D=10^{-3}/T$ (green crosses), whereas LNA again breaks down in
  this regime (blue crosses).

 While the LNA breaks down when the
distribution $P(\phi|t)$ becomes non-Gaussian as the coupling becomes
too weak, the PAM accurately describes $P(\phi|t)$ in the
low-coupling, low-noise regime, as it allows for non-Gaussian
distributions. However, the PAM does assume that $\phi(t)$ follows
$\omega t$. As a result it breaks down when the coupling becomes
large, causing the average input-output relation $\overline{\phi}(t)$ to
deviate markedly from $\omega t$, an effect that can be captured by
the LNA. PAM also breaks down when $\epsilon$ is small and $\omega
\approx \omega_0$, yet $D$ is large: now the large diffusion constant
causes the instantaneous $\phi(t)$ to deviate markedly from
$\omega t$. This regime can, however, be described by linear-response
theory.

\subsection{Linear response theory}
When the coupling strength is weak yet the diffusion constant is
large, $\phi(t)$ at any moment in time will tend to deviate strongly
from $\omega_0 t$, but the steady-state distribution will be close to
that of a noisy, free running oscillator, $P_0(\phi) = 1 /
(2\pi)$. The full distribution can then be obtained as a perturbation
to this distribution. This is the central idea of linear-response
theory (LRT).

We start with the Fokker-Planck
equation for the evolution of $P(\phi,t)$:
\begin{equation}\label{diff}
\partial_t P(\phi,t) = D \partial_{\phi}^2 P(\phi,t) + \omega_0 \partial_{\phi} P(\phi,t) + L(t) \partial_{\phi} \left[ Z(\phi) P(\phi,t) \right].
\end{equation}
We now consider the external signal $L(t) Z(\phi)$ to be a weak
perturbation of the free-running system. To this end, we rewrite the
above equation as:
\begin{equation}
  \partial_t P(\phi,t) =\left[\mathcal{F}_0 + \epsilon
    \mathcal{F}_1(t)\right] P(\phi,t) 
\elabel{FPF0F1}
\end{equation}
where $\mathcal{F}_0$ is the operator that defines the time evolution
of the unperterburted system and $\mathcal{F}_1$ that due to the perturbation:
\begin{align}
\mathcal{F}_0&=+D \partial^2_{\phi} +\omega_0 \partial_{\phi}\\
\mathcal{F}_1(t) &= +L(t) \partial_{\phi}Z(\phi) + L(t) Z(\phi) \partial_{\phi} 
\end{align}
Furthermore, we expand $P(\phi,t)$ as:
\begin{equation}
P(\phi,t) \simeq p_0(\phi,t) + \epsilon p_1(\phi,t) + \epsilon^2 p_2(\phi,t) + \mathcal{O}(\epsilon^3)
\end{equation}
Substituting this expression into \eref{FPF0F1}, and keeping only
terms up to order $\epsilon$, we find:
\begin{align}
&\mathcal{O}(0) \;\;\;\;\;\;\;\;\; \mathcal{F}_0  p_0(\phi,t)= \partial_t p_0(\phi,t) \elabel{FPp0}\\
&\mathcal{O}(\epsilon) \;\;\;\;\;\;\;\;\; \partial_t p_1(\phi,t) -\mathcal{F}_0  p_1(\phi,t)   = \mathcal{F}_1  p_0(\phi,t)\elabel{FPp1}
\end{align}

We are interested in the solutions that satisfy the periodic boundary
conditions:
\begin{align}
p_i(\phi,t) &= p_i(\phi+2\pi,t)  \elabel{BCphi}\\
 \partial_{\phi} p_i(\phi,t) & =  \partial_{\phi}\elabel{BCdphi}
 p_i(\phi+2\pi,t),
\end{align}
for both $i=0,1$. Moreover, in steady state, for $t\to \infty$, it must hold that
\begin{align}
p_i(\phi,t) &= p_i (\phi,t+T).\elabel{BCt}
\end{align}

\eref{FPp0} describes the diffusion of a particle
  with drift.  The steady-state solution, which obeys \erefsrange{BCphi}{BCt}, is
\begin{equation}
\lim_{t\rightarrow \infty} p_0(\phi,t) = \frac{1}{2\pi}
\elabel{p0ss}.
\end{equation}
 Clearly,
$p_0(\phi,t)$ in steady state is flat, which means that any deviation
in the steady-state solution for $P(\phi,t)$ from the flat distribution
must be contained in $p_1(\phi,t)$. Since $p_1(\phi,t)$ is, by
construction, a small perturbation, this approach will be 
accurate only when the full distribution is sufficiently flat, which means
that the diffusion constant cannot be too small.

To obtain $p_1(\phi,t)$, we proceed by  substituting the
solution for $p_0 (\phi,t)$, \eref{p0ss}, into \eref{FPp1}, yielding
\begin{align}
\partial_t p_1(\phi,t) -D \partial_{\phi}^2 p_1(\phi,t)
-\omega_0 \partial_\phi p_1(\phi,t)  =& L(t)  p_0(\phi,t) \partial_{\phi}Z(\phi).
\end{align}
The solution to this non-homogeneous heat equation is given by
\begin{align}
p_1(\phi,t) = &\int_0^{2\pi} d \xi G(\phi-\omega_0 t,\xi,t)
f(\xi)\nonumber\\ + &\int_0^{2\pi} \int_0^t d\tau d \xi
G(\phi-\omega_0 t,\xi,t-\tau) A(\xi,\tau),\elabel{solp1}
\end{align}
where $f(\phi)$ is the initial condition, $G(\phi-\omega_0 t,\phi_0,t,t_0)$ is
the Green's function of the unperturbed diffusion operator with drift, and
 $A(\phi,t) \equiv L(t)
p_0(\phi,t) \partial_{\phi}Z(\phi)$. This expression holds for any
$t$, not only for the steady-state solution.

To obtain the steady-state solution, we aim to find the initial
condition $P(\phi,t) = f(\phi)$ that folds back onto it self after a
time $T$: $P(\phi,t+T) = P(\phi,t) = f(\phi)$. To this end, we
evaluate \eref{solp1} for $t=T$, to arrive at the Fredholm equation of
the second kind:
\begin{equation}
\elabel{fred}
f(\phi) = \int_0^{2\pi} d\xi f(\xi) G(\phi,\xi,t=T) + Q(\phi)
\end{equation}
where $Q(\phi)$ is given by \eref{Q}. The above equation can be solved
analytically, see Appendix \ref{sec:LRT}.

\fref{KLtest} and \fref{MutInfoOpt} show, respectively, that the LRT accurately
describes $P(\phi,t)$ and hence the mutual information in the regime
that the coupling is weak and the diffusion constant is large. In
contrast to the phase-averaging method, the LRT breaks down for
smaller diffusion constant. The reason is that then $P(\phi,t)$
deviates increasingly from the uniform distribution, $p_0(\phi,t)=1/(2\pi)$,
and the full solution $P(\phi,t)$ can no longer be treated as a weak
perturbation to $p_0$.

\section{Discussion}
\label{sec-discussion}
The phase-response curves that have been measured experimentally often
have a positive lobe and a negative one, separated by a deadzone where the
coupling strength is zero \cite{Pfeuty2011}. However, the width of the
deadzone varies considerably from organism to organism. Here, we
asked how the optimal phase-response curve depends on the intrinsic noise
in the system, using the mutual information as a performance measure.

Information theory predicts that the number of signals that can be
transmitted reliably through a communication channel depends on the
shape of the input distribution, the input-output relation, and the
noise in the system. These arguments apply to any signaling system and
the circadian clock is no exception. 

When the input distribution is flat and the noise is low, then, in
general, the optimal input-output relation is linear. The
phase-oscillator model of the clock obeys this rule: the input
distribution $p(t) = 1/T$ is flat, and the optimal input-output
relation $\phi(t)$ is indeed linear in the low-noise regime
(\fref{TradeOff}B,C). Such a linear input-output relation is obtained
for an intrinsic period that is close to 24 hrs and for a deadzone
that is relatively large (\frefstwo{GlobOpt}{DeltaPhi12}). Our analysis thus predicts
that less-noisy circadian clocks exhibit a relatively large
deadzone. Interestingly, the rule also explains why for a constant
deadzone, in the low-noise limit, the optimal intrinsic frequency
decreases as the coupling strength increases (see \fref{TradeOff}A).

In the large-noise regime, containment of noise becomes
paramount. This inivetably requires a large coupling strength. While a
strong coupling distorts the input-output relation, which tends to
reduce information transmission, it also reduces the noise, enhancing
information transmission (\fref{TradeOff}B,C). The stability is
further enhanced by increasing the intrinsic frequency and reducing
the width of the deadzone (\fref{DeltaPhi12}). Indeed, our results
predict that noisy circadian systems feature a smaller deadzone and a
higher intrinsic frequency.

These results have been obtained by reducing the circadian clock to a
phase-oscillator model. It is useful to briefly review the generality
and limitations of this approach. The mutual information obeys
$I({\bf n};t)\geq I(R,\phi;t)\geq
I(\phi;t)$. Hence, any mapping of ${\bf n}$ to $\phi$ makes it
possible to put a lower bound on the mutual information. The
bound will be tight when the phase, according to this mapping,
contains most of the information on time. 

Another question is whether the model that we use to describe the
evolution of the phase is accurate. Phase-oscillator models have
commonly been employed to describe oscillatory systems, yet they are
typically described as being valid in the limits of weak driving and
low noise: this ensures that the coupled system stays close to the
limit cycle of the unperturbed, deterministic system, so that the
coupling function and the diffusion constant can be approximated by
their values on that limit cycle \cite{Pikovsky2003}. Here, having
derived the phase oscillator description in the weak coupling limit,
we then proceed to study it for arbitrary values of $\epsilon$ and
$D$.  This might at first glance seem self-contradictory. It should be
realized, however, that biochemical noise and coupling can have two
distinct effects: they can affect the dynamics {\em along} the limit
cycle, i.e. of $\phi$, and/or they can cause the system to move {\em
  away} from the limit cycle. Only perturbations in the latter
direction, orthogonal to the limit cycle, need be small for the phase
oscillator description to apply.  Moreover, $\epsilon$ and $D$ are
dimensionful parameters that can only be meaningfully be said to be
large or small in comparison to another parameter, and the appropriate
parameter for comparison is different for perturbations along and
orthogonal to the limit cycle.  Thus, it is entirely possible for
$\epsilon$ and $D$ to be small compared to the rate of relaxation to
the limit cycle, implying that neither the external driving nor the
noise can force the system far from the limit cycle and that the phase
oscillator model is a good approximation, but simultaneously for one
or both of $\epsilon$ and $D$ to be large compared to $\omega_0$, so
that perturbations to the phase dynamics are not weak.  We imagine
that just such a situation holds here: $D$ and $\epsilon$ can become
bigger than $\omega_0$---meaning that the noise and the coupling can
induce large changes in $\phi$---but, even for large $D/\omega_0$ and
$\epsilon/\omega_0$, the system in our model does not significantly
move off the limit cycle. It remains an open question how for a given, particular clock biochemical noise and strong coupling to
an entrainment signal affect the dynamics: how far does the system
move away from its limit cycle, and how much do the diffusion constant
and the coupling function then change?  The detailed and minimal
biochemical network models that have been developed for the
cyanobacterium {\it Synechococcus elongatus} would make it possible to
investigate this question in detail
\cite{VanZon2007,Rust2007,Zwicker2010,Phong2012,Paijmans:2016fd,Paijmans:2017gx,Paijmans:2017jm,Paijmans:2016uh}

  Our work shows that the behavior of the coupled phase oscillator can
  be accurately described by three different theories, which each work
  best in a different parameter regime. In the regime of weak
  coupling, low noise, and intrinsic frequency close to the driving
  frequency, the phase-averaging method is very accurate. In the
  regime that the driving is strong compared to the diffusion
  constant, the linear-noise approximation is most accurate. These are
  the two most relevant regimes for understanding the design of
  circadian clocks. There is also another regime, however, namely that
  of weak coupling and high noise, and in this regime linear-response
  theory is very accurate. That linear-response theory can
    describe any regime at all is perhaps surprising, since it has
    been argued that this theory should be applied to phase oscillators only with the greatest care
  \cite{Pikovsky2003}. The argument is that small but resonant
    forcing can have effects on $\phi$ that build up over time,
    meaning that the effect of perturbations that are nominally of order $\epsilon$, and thus small, will eventually
    become large with time. However, noise can pre-empt this accumulation of
    resonant perturbations by effectively randomizing the phase and erasing the memory of earlier perturbations before they are able to accumulate over time. As a result, the full distribution
    of the phase can be written as a small perturbation around the
    uniform distribution, and this does make it possible to apply
    linear-response theory. While this regime is probably less
    relevant for understanding biological clocks, this approach may be
    useful in other contexts.

Finally, we have focused on the optimal design of the clock as a
function of the intrinsic noise in the system. As Pfeuty {\it et al.}
have shown, fluctuations in the input signal are an important
consideration for understanding the design of circadian clocks
\cite{Pfeuty2011}. It will be interesting to see whether maximizing
the mutual information will reveal new design principles for clocks
driven by fluctuating signals.

\emph{Acknowledgements}: We thank Jeroen van Zon for a critical reading
  of the manuscript. This work is supported by the research programme of the
  Foundation for Fundamental Research on Matter (FOM), which is part
  of the Netherlands Organisation for Scientific Research
  (NWO), and by NSF grant DMR-1056456 (DKL).

\appendix
\section{Arnold tongue of the deterministic model}
For completeness, we give here the inequalities for all scenarios.
{\bf Scenario 1:} As discussed in the main text: $\phi_3 - 2\pi < \phi_s <
\phi_1$; $t_2 < T/2 < t_3$. If $\epsilon_- \leq \omega_0$, then

\begin{align}
T &\leq \frac{2\pi - \epsilon_-\Delta
  \phi_{12}/\omega_0}{\omega_0 - \epsilon_-/2}\\ 
T &> \frac{2\pi + \epsilon_+\Delta \phi_{12}
  /\omega_0}{\epsilon_+/2+\omega_0}\\
T& <\frac{2\pi + \epsilon_+\Delta \phi_{12}
  /\omega_0 + \Delta \phi_{23}(\epsilon_++\epsilon_-)/(\omega_0 - \epsilon_-)}{\epsilon_+/2+ \omega_0}\\
T&>\frac{(\Delta
  \phi_{13}-2\pi)(\epsilon_++\epsilon_-)/(\omega_0+\epsilon_+) + 2 \pi  - \epsilon_-\Delta
  \phi_{12}/\omega_0}{\omega_0 - \epsilon_-/2}
\end{align}
If $\epsilon_- > \omega_0$ then
\begin{align}
T& \leq \frac{2 (2\pi - \Delta \phi_{12})}{\omega_0}\\
T&> \frac{2 \pi - \Delta \phi_{12} + (\Delta
  \phi_{12}/\omega_0)(\epsilon_++\omega_0)}{\epsilon_+/2 + \omega_0}\\
T&>\frac{2\Delta \phi_{23}}{\omega_0}
\end{align}

{\bf Scenario 2:} $\phi_1 < \phi_s < \phi_2$; $0< t_2 < T/2 < t_3 < t_1 <
T$.  For
$\epsilon_- < \omega_0$, the evolution of $\phi(t)$ is given by
\begin{align}
\phi_s + \omega_0 t_2 + (-\epsilon_- + \omega_0) (T/2-t_2) + \omega_0
T/2 = \phi_s + 2 \pi.
\end{align}
This yields:
\begin{align}
t_2 &= \frac{2\pi - T (\omega_0 - \epsilon_-/2)}{\epsilon_-} < T /2\,\,\&\,>0\\
t_3 &=\frac{\Delta\phi_{23}}{\omega_0 - \epsilon_-} + t_2  > T / 2\\
t_1 &= t_2 - \Delta \phi_{12}/\omega_0 + T < T.\\
\phi_s &= \phi_2 - \omega_0 t_2 > \phi_1.
\end{align}
This yields the following inequalities:
\begin{align}
T&> \frac{2\pi}{\omega_0}\\
T&< \frac{2\pi}{\omega_0 - \epsilon_-/2}\\
T&<\frac{2\pi + \Delta \phi_{13} \epsilon_-}{\omega_0-\epsilon_-/2}\\
T&>\frac{2\pi - \epsilon_- \Delta \phi_{12}/\omega_0}{\omega_0 - \epsilon_-/2}
\end{align}
If $\epsilon_- > \omega_0$, the equation to solve is
\begin{align}
\phi_s + \omega_0 t_2 + \omega_0 T / 2 = \phi_s + 2 \pi.
\end{align}
The solution is
\begin{align}
t_2 &= \frac{2 \pi}{\omega_0} - T / 2 < T /2 \\
t_3 &= \infty > T / 2.\\
t_1 &= t_2 - \Delta \phi_{12}/\omega_0 + T < T.\\
\phi_s &= \phi_2 - \omega_0 t_2 > \phi_1 \,\,\&\, < \phi_2.
\end{align}
This yields the following inequalities
\begin{align}
T &> 2 \pi / \omega_0\\
T &> \frac{2(2\pi - \Delta \phi_{12})}{\omega_0}\\
T & < 4\pi / \omega_0
\end{align}
This scenario is stable, because $\phi(t)$ between $t=0$ and $t=t_2$
is steeper than $\phi(t)$ between $t_2$ and $T/2$.

{\bf Scenario 3:} $\phi_2 < \phi_s < \phi_3$; $0< t_2 < T / 2$.
If $\epsilon_- < \omega_0$ then
\begin{align}
\phi_s + (-\epsilon_- + \omega_0) T/2 + \omega_0 T / 2 = \phi_s + 2
\pi.
\end{align}
This equation does not depend on $t_i$. There is only one period that
fits the solution:
\begin{align}
T &= \frac{2\pi}{\omega_0 - \epsilon_-/2}.
\end{align}
This period is on the boundary of the Arnold tongue of scenario
2. This solution seems degenerate, being neither stable nor unstable. 

If $\epsilon_- > \omega_0$, the equation that solves $\phi(t)$ is
\begin{align}
\phi_s + (-\epsilon_- + \omega_0) t_2 + \omega_0 T/2 = \phi_s + 2 \pi.
\end{align}
The solution is
\begin{align}
t_2 &= \frac{2\pi - \omega_0 T / 2}{-\epsilon_- + \omega_0}\elabel{sc3_t2}
\\
\phi_s &= \phi_2 + \omega_0 T / 2 - 2\pi.
\end{align}
The requirement that $t_2 > 0$, yields the inequality
\begin{align}
T &> \frac{4 \pi}{\omega_0},
\end{align}
because the denominator of \eref{sc3_t2} is negative. The requirement
that $t_2 < T / 2$ yields 
\begin{align}
\frac{2\pi - T(\omega_0-\epsM/2)}{\omega_0 - \epsM} < 0.
\end{align}
Since the denominator is negative for $\epsM>\omega_0$, this means that $(2\pi - T(\omega_0
- \epsM/2))>0$. When $\epsM > 2\omega_0$, this is true for any $T$. When
$\epsM < 2 \omega_0$ (but still larger than $\omega_0$ because
otherwise there is no solution at all, see above), then
\begin{align}
T &< \frac{2\pi}{\omega_0 - \epsilon_-/2}.
\end{align}

The constraints $\phi_2 < \phi_s < \phi_3$ yield
\begin{align}
T&> \frac{4\pi}{\omega_0}\\
T&< \frac{2(\Delta \phi_{23}+2\pi)}{\omega_0}.
\end{align}
This solution is rather strange. When the light comes up, the clock is
being driven backwards. The solution seems stable, though. In fact, it
seems extremely stable: after one period, the system is back on its
limit cycle.

{\bf Scenario 4:} $\phi_3 - 2 \pi < \phi_s < \phi_1$; $0<t_1 < T/2 < t_2$.
The equation that determines the steady state is
\begin{align}
\phi_s + (\omega_0 + \epsilon_+) t_1 + \omega_0 (T/2 - t_1) + \omega_0
T / 2 = \phi_s + 2 \pi.
\end{align}
The solution is
\begin{align}
t_1 &= \frac{2\pi - \omega_0 T}{\epsilon_+} < T/2 \,\,\&\, > 0\\
t_2 &= t_1 +\frac{\Delta \phi_{12}}{\omega_0} > T/2\\
\phi_s &= \phi_1 - (\epsilon_+ + \omega_0) t_1 = \phi_1 -
(\epsilon_+ + \omega_0)(2\pi - \omega_0 T)/\epsilon_+
\end{align}
The conditions for $T$ are
\begin{align}
T&\leq \frac{2\pi}{\omega_0}\\
T&>\frac{2\pi}{\omega_0 + \epsilon_+/2}\\
T&<\frac{2\pi+\Delta \phi_{12}\epsilon_+/\omega_0}{\epsilon_+/2+\omega_0}\\
T&>\frac{\epsilon_+\Delta \phi_{13} + 2 \pi \omega_0}{\omega_0
  (\omega_0 + \epsilon_+)}.
\end{align}

{\bf Scenario 5:} $\phi_3  - 2 \pi < \phi_s < \phi_1$; $t_1 > T/2$. The
governing equation is 
\begin{align}
\phi_s + (\epsilon_+ + \omega_0) T/2 + \omega_0 T / 2 = \phi_s + 2 \pi.
\end{align}
This means that
\begin{align}
T&= \frac{2\pi}{\omega_0 + \epsilon_+/2}.
\end{align}
Clearly, for each $\epsilon_+$ there is only one period, not a range
of periods. Since $\phi(T/2) = \phi_s + (\epsilon_+ + \omega) T /2$,
which must be smaller than $\phi_1$, and $\phi_s > \phi_3 - 2 \pi$, we
find that there exists only a solution if $\Delta \phi_{13} < 2 \pi
\omega_0 / (\epsilon_+ + 2 \omega_0)$. Hence, for given $\phi_1$ and
$\phi_3$, this puts an upper bound on $\epsilon_+$. If a solution
exists, the starting phase $\phi_s$, must lie in the range $\phi_3 - 2
\pi< \phi_s < \phi_1 - \pi (\epsilon_+ + \omega_0) / (\epsilon_+ / 2 +
\omega_0)$. Moreover, the solution is neutral; it does not relax back
to a unique $\phi_s$. In fact, this is a very general observation: if
the solution is neutral, it means that there can only be locking for
one value of the period. Being able to locking over a range of periods
of the driving signal, means that the clock should be able to adjust
its period by changing the phase; but a neutral solution means that
changing the phase does not lead to a change in its period.

{\bf Scenario 6:} $\phi_3 - 2 \pi < \phi_s < \phi_1$; $0<t_1<t_2<t_3 <
T/2$. This scenario can only arise when $\epsilon_- < \omega_0$,
because otherwise the system never makes it to $\phi_3$ before the sun
sets. 
The equation to be solved is then:
\begin{align}
\phi_s + (\epsilon_++\omega_0) t_1 + \Delta \phi_{13} +
(\omega_0+\epsilon_+) (T/2 - t_3) + \omega_0 T / 2 \nonumber\\
 = \phi_s + 2 \pi.
\end{align}
This equation can be solved by noting that $\Delta \phi_{12} =
\omega_0 (t_2 - t_1)$ and $\Delta \phi_{23} = (-\epsilon_- + \omega_0)
(t_3 - t_2)$. It follows that there is only one period that satisfies
the above equation:
\begin{align}
T = \frac{2\pi -\Delta \phi_{23}+\epsilon_+ \Delta \phi_{12}/\omega_0 +
  (\epsilon_+ + \omega_0) \Delta
  \phi_{23}/(-\epsilon_-+\omega_0)}{\omega_0 + \epsilon_+/2}
\end{align}
Clearly, for a given $\epsilon_-$ and $\epsilon_+$ there is only one
period, not a range of periods to which the system can entrain. This
means that the solution is neutral, which can indeed be understood by
noting that the initial slope at $t=0$, $\omega_0+\epsilon_+$, is the
same as that $t=T/2$.
The condition for the solution to exist is that $\phi(T/2) = 2 \pi +
\phi_s - \omega_0 T / 2 > \phi_3$. This yields for $\phi_s$:
\begin{align}
\phi_3 - 2 \pi + \omega_0 T / 2 < \phi_s < \phi_1.
\end{align}
There is thus only a solution when
\begin{align}
T < \frac{2(2\pi - \Delta \phi_{13})}{\omega_0}.
\end{align}
One could use this condition to determine the range of
$\epsilon_{+/-}$ over which there is a solution, given $\phi_1,\phi_2,
\phi_3$. But since this scenario only yields one line in the phase
diagram, we do not pursue this further. 

{\bf Scenario 7:} $\phi_1 < \phi_s < \phi_2$; $0< T / 2 < t_2 < t_3 <
t_1$. The governing equation is
\begin{align}
\phi_s + \omega_0 T / 2  + \omega_0 T / 2 = \phi_s + 2 \pi.
\end{align}
This indeed yields only one solution
\begin{align}
T = \frac{2\pi}{\omega_0}.
\end{align}
Indeed, there only exists a solution when the driving frequency equals
the intrinsic frequency, which is to be expected, since with this
solution the system does not see the driving. The solution exists only
if $\Delta\phi_{12} > \pi$. This solution is
neutral, in that all solutions $\phi_1 < \phi_s < \phi_2$ are valid,
for all values of $\epsilon_{-/+}$. One may wonder what that implies
for the dynamics. If one would perform a simulation for
$\epsilon_{-/+}>0$ and $\omega=\omega_0$, and if one would then start
with $\phi_1 < \phi_s < \phi_2$, then due to the noise the simulation
would initially perform a random walk where initially, at the
beginning of each day, the phase of the clock would fluctuate between
$\phi_1$ and $\phi_2$. However, once the oscillator due to noise would
cross the boundary $\phi_1$, then the system will be driven to a
solution that is described under scenario 1.

{\bf Scenario 8:} $\phi_1 < \phi_s < \phi_2$; $0<t_2 < t_3 < T_2 <
t_1$. There can only be a solution, if it exists, when $\epsilon_- <
\omega_0$. For $\epsM > \omega_0$ the system never makes it to
$\phi_3$ before $T/2$. The governing equation is 
\begin{align}
\phi_s + \omega_0 t_2 + \Delta \phi_{23} + (\epsilon_+ + \omega_0)
(T/2 - t_3) + \omega_0 T / 2 = \phi_s + 2 \pi.
\end{align}
To solve this, we note that
\begin{align}
t_3 = t_2 + \Delta \phi_{23} / (-\epsilon_-+\omega_0).
\end{align}
This yields:
\begin{align}
t_2 = \frac{T(\omega_0 + \epsilon_+/2) - 2 \pi - \Delta
\phi_{23}(\epsilon_++\epsilon_-) / (\omega_0 - \epsilon_-)}{\epsilon_+}.
\end{align}
We further have
\begin{align}
\phi_s = \phi_2 - \omega_0 t_2.
\end{align}
The condition $t_2>0$ yields
\begin{align}
T > \frac{2\pi + \Delta \phi_{23}(\epsilon_+ + \epsilon_-)/(\omega_0 -
  \epsilon_-)}{\omega_0 + \epsilon_+/2}
\end{align}
The condition $t_3 < T/2$ yields
\begin{align}
T<\frac{2
  \pi+\Delta\phi_{23}\epsilon_-/(\omega_0 - \epsilon_-) }{\omega_0}.
\end{align}
The condition $\phi_1 < \phi_s = \phi_2 - \omega_0 t_2$ yields
\begin{align}
T < \frac{2 \pi + \epsP \Delta \phi_{12}/\omega_0 + \Delta \phi_{23}
  (\epsP + \epsM)/(\omega_0 - \epsM)}{\omega_0 + \epsP/2}.
\end{align}

The Arnold tongue of this scenario is embedded in those of scenarios
1 and 2. The solution corresponding to this scenario is indeed
unstable: the system either converges to the solution of scenario 1 or
2. This can be easily proven by noting that the time it takes to cross
$\Delta \phi_{23}$ is constant, as is the time to cross the night. The
change in the phase a period later is then the change in the phase at
$\phi(T/2)$. This is given by $\delta \phi(T/2) = \partial \phi(T/2)
/\partial t_3 \delta t_1 =  \partial \phi(T/2)
/\partial t_3 \delta \phi_s / \omega_0 = (\epsP + \omega_0) /
\omega_0 \delta \phi_s$, where we have noted that
$\delta t_1 = - \delta \phi_s / \omega_0$ and $\partial
\phi(T/2) /\partial t_3 = -(\epsP + \omega_0)$. Because  $(\epsP + \omega_0) /
\omega_0 > 1$, the change in the phase after a full period is larger
than the initial change in the phase:
$\delta \phi(T) = \delta \phi(T/2) > \delta \phi_s$. The solution is unstable. 

{\bf Scenario 9:} $\phi_1 < \phi_s < \phi_2$; $t_2 < t_3 < t_1 <
T_2$. There can only be a solution if $\epsM < \omega_0$. The equation to be solved is
\begin{align}
\phi_s + \omega_0 t_2 + 2\pi - \Delta \phi_{12} + \omega_0 (T-t_1) =
\phi_s + 2\pi,
\end{align}
which gives
\begin{align}
T =  \Delta \phi_{12}/\omega_0 + t_1 - t_2.
\elabel{Tt1t2scen9}
\end{align}
We further have
\begin{align}
t_1 - t_2 = \frac{2\pi-\Delta \phi_{13}}{\omega_0+\epsP}+\frac{\Delta
  \phi_{23}}{\omega_0-\epsM}.
\end{align}
Hence, 
\begin{align}
T =  \frac{\Delta \phi_{12}}{\omega_0} +  \frac{2\pi-\Delta \phi_{13}}{\omega_0+\epsP}+\frac{\Delta
  \phi_{23}}{\omega_0-\epsM},
\elabel{Tscen9}
\end{align}
which we could have written down right away upon somewhat more careful
thinking. We can obtain a bound on the parameters that allow a
solution by noting that $0<t_1 - t_2 < T/2$. Combining with
\eref{Tt1t2scen9} yields $\Delta \phi_{12}/\omega_0 < T < 2\Delta
\phi_{12}/\omega_0$. Combing this with \eref{Tscen9} yields
\begin{align}
 \frac{\Delta \phi_{12}}{\omega_0} < \frac{2\pi-\Delta \phi_{13}}{\omega_0+\epsP}+\frac{\Delta
  \phi_{23}}{\omega_0-\epsM}.
\end{align}
A visual inspection illustrates this containt very clearly. The
parameter $\epsM$ should be small, that is not close to
unity. A large $\epsP$ also helps.

{\bf Scenario 10:} $\phi_2 < \phi_s < \phi_3$; $0<T_2 < t_3, t_1,
t_2$.  Both for $\epsM<\omega_0$ and $\epsM>\omega_0$, the scenario
corresponds to that of scenario 3, but with $\epsM < \omega_0$) in that
scenario. There is only a solution for
\begin{align}
T= 2\pi / (\omega_0 -
\epsM/2).
\end{align}

{\bf Scenario 11:} $\phi_2 < \phi_s \phi_3$; $0<t_3 < T/2 < t_1,
t_2$. Only if $\epsilon_- < \omega_0$ may a solution exist: if $\epsM
> \omega_0$, we are back to scenario 3 or 10. The
governing equation is
\begin{align}
\phi_s + (-\epsM + \omega_0) t_3 + (\epsP + \omega_0) (T/2 - t_3) +\nonumber\\
\omega_0 T /2 = \phi_s + 2\pi.
\end{align}
The solution is 
\begin{align}
t_3 &= \frac{T (\omega_0 + \epsP/2)-2\pi}{\epsP + \epsM}\\
\phi_s &= \phi_3 - (\omega_0 - \epsM) t_3.
\end{align}
The condition $t_3 > 0$ yields
\begin{align}
T>\frac{2\pi}{\omega_0 + \epsP/2}.
\end{align}
The condition $t_3 < T / 2$ yields the inequality
\begin{align}
T < \frac{2\pi}{\omega_0 - \epsM/2}.
\end{align}
The condition $\phi_s > \phi_2$ yields
\begin{align}
T < \frac{\Delta \phi_{23}(\epsP+\epsM) / (\omega_0 - \epsM)  + 2 \pi}{\omega_0
  + \epsP/2}.
\end{align}
The condition $t_1 > T / 2$ yields the inequality
\begin{align}
T > \frac{2\pi - (2\pi - \Delta \phi_{13})(\epsP+\epsM)/(\epsP +
  \omega_0)}{\omega_0-\epsM/2}.
\end{align}

The solution space overlaps with those of scenarios 1 -
3. Interestingly, we find again that this solution is unstable:
$\delta \phi(T) = \delta \phi(T/2) = \partial \phi(T/2) / \partial t_3
\delta t_3 = -(\omega_0+\epsP) \delta t_3 = -(\omega_0 + \epsP)\partial
t_3 /\partial \phi_s \delta \phi_s = (\omega_0 + \epsP) / (\omega_0 -
\epsM) \delta \phi_s > \delta \phi_s$. We thus can see that when
$\phi(t)$ is convex for $0<t<T/2$, the solution tends to be unstable.  

{\bf Scenario 12:} $\phi_2 < \phi_s < \phi_3$; $t_3,t_1 < T/2 <
t_2$. Only if $\epsM < \omega_0$ may a solution exist. The governing
equation is
\begin{align}
\phi_s + (-\epsM + \omega_0) t_3 + (2\pi-\Delta \phi_{13}) + \omega_0
(T/2 - t_1) \nonumber \\
+ \omega_0 T/2 = \phi_s + 2\pi.
\end{align}
Exploiting that $t_1 = t_3 + (2 \pi - \Delta \phi_{13})/(\epsP +
\omega_0)$, the solution is
\begin{align}
t_3 &= \frac{\omega_0 T - \Delta \phi_{13} - \omega_0 (2\pi - \Delta
  \phi_{13})/(\epsP+\omega_0)}{\epsM}\\
\phi_s &= \phi_3 - (\omega_0 - \epsM) t_3.
\end{align}
The condition $t_3>0$ yields the inequality
\begin{align}
T>\frac{\Delta \phi_{13}}{\omega_0} + \frac{2\pi - \Delta
    \phi_{13}}{\epsP + \omega_0}.
\end{align}
The condition $t_1 < T/2$ gives
\begin{align}
T<\frac{\Delta \phi_{13}+ (\omega_0 - \epsM)(2\pi - \Delta \phi_{13})/(\epsP +
  \omega_0)}{\omega_0 - \epsM/2}.
\end{align}
The condition $t_2 = t_1 + \Delta \phi_{12}/\omega_0 > T/2$ yields
\begin{align}
T>\frac{\Delta \phi_{13}+ (\omega_0 - \epsM)(2\pi - \Delta \phi_{13})/(\epsP +
  \omega_0) - \epsM \Delta \phi_{12}/\omega_0}{\omega_0 - \epsM/2}.
\end{align}
The condition $\phi_s>\phi_2$ yields the inequality
\begin{align}
T < \frac{\Delta \phi_{13} + \omega_0 (2\pi - \Delta \phi_{13})/(\epsP
  + \omega_0) + \epsM \Delta \phi_{23}/(\omega_0 - \epsM)}{\omega_0}.
\end{align}

This curve is convex, that is the part of $\phi(t)$ that really
matters is convex: the initial slope near $t=0$, $\omega_0 - \epsM$,
is smaller than the slope near $t=T/2$, which is $\omega$. This gives
an unstable solution.

{\bf Scenario 13:} $\phi_2 < \phi_s < \phi_3$; $t_3,t_1,t_2 <
T/2$. Again, a solution may only exist if $\epsM < \omega_0$. The
central equation is
\begin{align}
\phi_s + (-\epsM + \omega_0) t_3 + (2\pi - \Delta \phi_{23}) +
(-\epsM+\omega_0) (T/2-t_2)\nonumber\\  + \omega_0 T / 2 = \phi_s + 2\pi.
\end{align}
The solution is 
\begin{align}
T = \frac{\Delta \phi_{23}}{\omega_0} + \frac{(\omega_0-\epsM)(t_2 -
  t_3)}{\omega_0}.
\end{align}
The time difference is
\begin{align}
t_2 - t_3 = \frac{\Delta \phi_{12}}{\omega_0} + \frac{2\pi - \Delta
    \phi_{13}}{\omega_0 + \epsP},
\end{align}
which gives for the period
\begin{align}
T = \frac{\Delta \phi_{23}}{\omega_0}+ \frac{\omega_0 -
    \epsM}{\omega_0} \left(\frac{\Delta \phi_{12}}{\omega_0} +
    \frac{2\pi - \Delta \phi_{13}}{\omega_0 + \epsP}\right).
\end{align}

\section{Heat maps mutual information as a function of coupling
  strength and intrinsic frequency}
\fref{MutInfo}A shows the mutual information as a function of the
coupling strength $\epsilon=\epsP=\epsM$ and intrinsic frequency
$\omega_0$, for one value of the diffusion constant,
$D=0.1/T$. \fref{MutInfoDs} shows the same plot, but then also for
$D=1/T$ and $D=10^{-4}/T$. For $D=10^{-4}/T$, the mutual information
shows very rich behavior, corresponding to intrincate locking behavior.

\begin{figure*}[t]
\centering
\includegraphics[width=18cm]{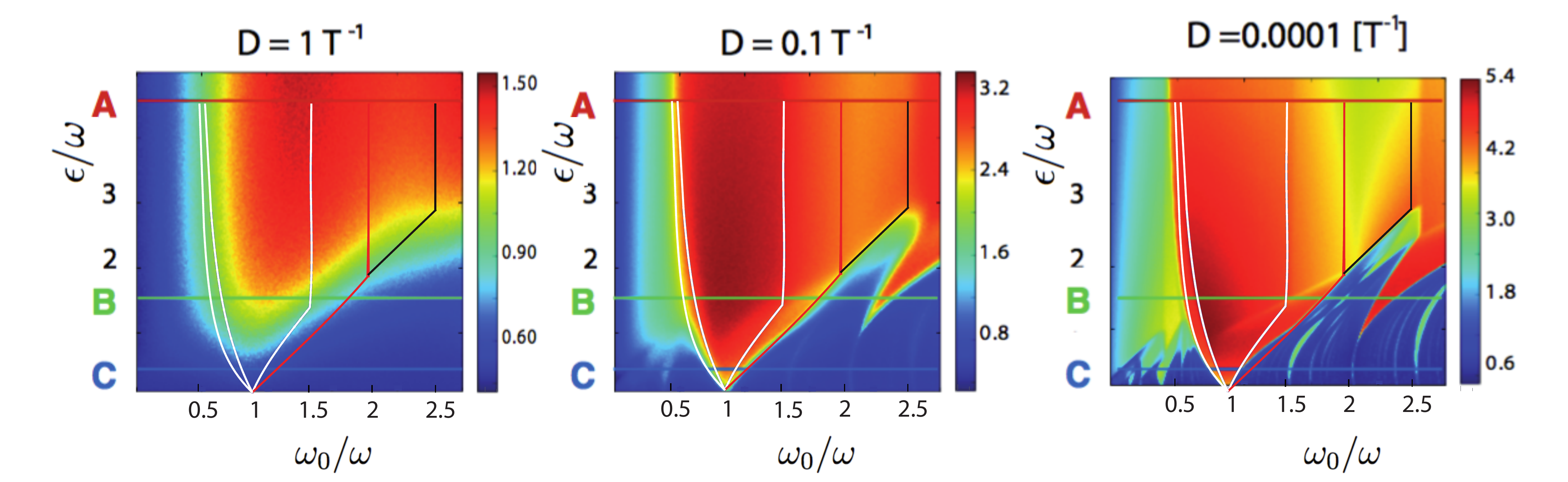}
\caption{The mutual information as a function of the coupling strength
  $\epsilon$ and the intrinsic frequency $\omega_0$, for three
  different values of the diffusion constant $D$. In all panels,
  $\Delta \phi_{12}=\Delta \phi_{23}=\pi/2$. Superimposed in black is
  the deterministic Arnold Tongue for scenarios 1 and 4. (A)
  $D=1/T$. (B) $D=0.1/T$ (the same panel as \fref{MutInfo}A). (C)
  $D=10^{-4}/T$. Note the rich behavior of the mutual information,
  corresponding to higher-order locking scenarios. }
\flabel{MutInfoDs}
\end{figure*}

\section{Linear-response theory}
\label{sec:LRT}
As shown in the main text, the evolution of $p_1(\phi,t)$ is given by
\begin{align}
\partial_t p_1(\phi,t) -D \partial_{\phi}^2 p_1(\phi,t)
-\omega_0 \partial_\phi p_1 (\phi,t)  = \nonumber\\
 L(t)  p_0(\phi,t) \partial_{\phi}Z(\phi).\elabel{app:eqp1}
\end{align}
The solution to this non-homogeneous heat equation is:
\begin{align}
p_1(\phi,t) = &\int_0^{2\pi} d \xi G(\phi-\omega_0 t,\xi,t)
f(\xi)\nonumber\\ + &\int_0^{2\pi} \int_0^t d\tau d \xi G(\phi
-\omega_0 t,\xi,t-\tau) A(\xi,\tau),\elabel{app:solp1}
\end{align}
where $f(\phi)$ is the initial condition, $G(\phi -\omega_0 t,\phi_0,t,t_0)$ is
the Green's function of the unperturbed diffusion operator, and
$A(\phi,t)\equiv L(t)
p_0(\phi,t) \partial_{\phi}Z(\phi) = L(t)/(2\pi) \left(-\delta (\phi-\phi_1)
- \delta (\phi-\phi_2) + 2\delta (\phi - \phi_3)\right)$. 

The Green's function is given by
\begin{align}
G(\phi-\omega_0 t,\phi_0,t)=& \sum_{j=0}^{\infty} e^{-j^2Dt}[A_j(\phi_0)
\cos(j(\phi -\omega_0 t)) \nonumber\\
&+ B_j(\phi_0)\sin(j(\phi - \omega_0 t))],
\end{align}
with
\begin{align}
&A_j (\phi_0)= \frac{1}{\pi} \int d \phi \delta(\phi - \omega_0
t-\phi_0) \cos (j(\phi-\omega_0 t)) =\frac{1}{\pi} \cos j \phi_0 \\
&B_j (\phi_0)= \frac{1}{\pi} \int d \phi \delta(\phi - \omega_0 t-\phi_0) \sin (j\phi') = \frac{1}{\pi} \sin j \phi_0\\
&A_0 =\frac{1}{2\pi}\\
&B_0=0
\end{align}
This yields:
\begin{align}
\label{Gre1}
G(\phi,\phi_0,t)=& \frac{1}{2\pi} + \frac{1}{\pi}\sum_{j=1}^{\infty}
e^{-i^2Dt} \times \nonumber\\
&\left[\cos(j\phi_0) \cos(j(\phi-\omega_0 t)) + \sin(j\phi_0)\sin(j(\phi -\omega_0 t))\right]
\end{align}
Substituting this expression into \eref{app:eqp1} and \eref{app:solp1}
gives
\begin{align}
p_1(\phi,t) &= \int_0^{2\pi} d \xi G(\phi,\xi,t) f(\xi) +  
\int_0^{2\pi} \int_0^t d\tau d \xi G(\phi,\xi,t-\tau) \times \nonumber\\
&\frac{L(\tau)}{2\pi} \left[-\delta(\xi- \phi_1) - \delta(\xi-\phi_2) + 2\delta(\xi-\phi_3) \right]\\
&= G_0(\phi,t) + \frac{1}{2\pi}\int_0^td \tau L(\tau) \Delta G(\phi,t-\tau),\elabel{dG}
\end{align}
where
\begin{align}
G_0(\phi,t) &=\int_0^{2\pi} d \xi G(\phi,\xi,t)f(\xi)\nonumber\\
\Delta G(\phi,t-\tau) &=  -G(\phi,\phi_1,t-\tau) -
G(\phi,\phi_2,t-\tau) \nonumber \\
&+ 2 G(\phi,\phi_3,t-\tau).
\end{align}
We can integrate the second term of \eref{dG} by parts. Calling the primitive of $\Delta G$,
\begin{align}
C(\phi,\tau;t) = \int d\tau \Delta G(\phi,t-\tau),
\end{align}
we find
\begin{align}
p_1(\phi,t) = G_0 (\phi,t) + [L(\tau)C(\phi,\tau;t)]_{\tau = 0}^{\tau = t}
- \int_0^t d\tau \frac{dL(\tau)}{d\tau} C(\phi,\tau;t).
\end{align}
Since $L(\tau)$ is a sequence of step functions,
\begin{align}
\frac{d L(\tau)}{d\tau} = \sum_{n=0}^{\infty} \delta(\tau-nT) -\delta(\tau-(nT+T/2)),
\end{align} 
which yields
\begin{align}
p_1(\phi,t) &= G_0 (\phi,t) + [L(\tau) C(\phi,\tau;t)]_0^t - \nonumber\\
& \sum_{n=0}^{nT<t}[ C(\phi,nT;t) - C(\phi,nT+T/2;t)]
\end{align}
\eref{app:eqp1} was derived assuming that the system is in steady
state, and $p(\phi,t)=p(\phi,t+T)$. This means that we only have to
consider times $0< t < T$, in which case only the first two terms
in the last sum on the right-hand side remain. More specifically, in steady state, the initial condition $f(\phi)$
equals the steady-state distribution, and $f(\phi) =
p(\phi,t=0)=p(\phi,t=T)$, meaning that the above expression reduces to
\begin{align}
f(\phi) &= G_0 (\phi,T) + Q(\phi)\nonumber \\
&=\int_0^{2\pi} f(\xi) G(\phi,\xi,t=T)  + Q(\phi),\elabel{app:fred}
\end{align}
where $Q(\phi)$ is defined as
\begin{align}
Q(\phi)\equiv&    - 2C(\phi,\tau=0;T) + C(\phi,\tau=T;T)
+\nonumber\\
& C(\phi,\tau=T/2;T)\elabel{Q}.
\end{align}

\eref{app:fred} an integral equation, more specifically an Fredholm equation
of the second type. The integration kernel $G(\phi,\xi,T)$ has the
form
\begin{align}
G(\phi,\xi,T) = \frac{1}{2\pi} + 
\frac{1}{\pi}\sum_{j=1}^{\infty} e^{-j^2DT} [\cos(j(\phi-\omega_0 T))\cos(j\xi) \nonumber\\+ \sin(j(\phi-\omega_0 T)) \sin(j\xi)].
\end{align}
We define  $G^*(\phi,\xi) = G(\phi,\xi)-1/(2\pi)$, and rewrite
\eref{app:fred} as:
\begin{align}
f(\phi) &= \int_0^{2\pi} d\xi f(\xi) G^*(\phi,\xi,t=T) + \frac{1}{2\pi}\int_0^{2\pi} d\xi f(\xi) + Q(\phi)\\
 &= \int_0^{2\pi} d\xi f(\xi) G^*(\phi,\xi,t=T) + \frac{1}{2\pi}  + Q(\phi)\\
 &= \int_0^{2\pi} d\xi f(\xi) G^*(\phi,\xi,t=T)  + Q^*(\phi),
\end{align}
where in going from the first to the second line we have exploited
that $f(\phi)$ is normalized, and in the last line we have defined
$Q^*(\phi) \equiv Q(\phi) - 1/(2\pi)$. The kernel $G^*(\phi,\xi,T)$ is
separable, and we can rewrite \eref{app:fred} as
\begin{align}
f(\phi) = \sum_{j=1}^{\infty} e^{-j^2DT}\int_0^{2\pi} d\xi f(\xi)
[\cos(j(\phi-\omega_0 T))\cos(j\xi)\nonumber\\
 + \sin(j(\phi-\omega_0 T)) \sin(j\xi)]  + Q_j^*(\phi)
\end{align}
with $Q^*(\phi) = \sum_j Q^*_j (\phi)$. 

To solve this integral equation, we define
\begin{align}
c1_j &\equiv \int_0^{2\pi} d\xi e^{-j^2D T} f(\xi) \cos(j\xi) \\
c2_j &\equiv \int_0^{2\pi} d\xi e^{-j^2D T} f(\xi) \sin(j\xi),
\end{align}
so that
\begin{align}
f(\phi) = \sum_j  \left[\cos(j(\phi-\omega_0 T)) c1_j +
  \sin(j(\phi-\omega_0 T)) c2_j\right. \nonumber\\
 \left.+ Q^*_j(\phi)\right].\elabel{app:f}
\end{align}
We now multiply both sides, once with $e^{-j^2 D T} \cos(j \phi)$ and
once with $e^{-j^2 DT}\sin(j \phi)$, and integrate from $0$ to
$2\pi$. On the left-hand side, this gives $c_{1j}$ and $c_{2j}$,
respectively. We then arrive at the following set of linear equations:
\begin{align}
c_{1j} &= \sum_k A_{jk} c_{1k} + B_{jk} c_{2k} + Q^*_{1k},\elabel{c1}\\
c_{2j}&=\sum_k C_{jk} c_{1k} + D_{jk} c_{2k} + Q^*_{2k},\elabel{c2}\\
\end{align} 
where
\begin{align}
A_{jk}&=\int_0^{2\pi} d\phi e^{-j^2D T}\cos(j\phi)\cos(k(\phi-\omega_0 T)) \\
B_{jk}&=\int_0^{2\pi} d\phi e^{-j^2D T}\cos(j\phi)\sin(k(\phi-\omega_0 T))\\
C_{jk}&= \int_0^{2\pi} d\phi e^{-j^2D T}\sin(j\phi)\cos(k(\phi-\omega_0 T)) \\
D_{jk}&=\int_0^{2\pi} d\phi e^{-j^2D T}\sin(j\phi)\sin(k(\phi-\omega_0 T))\\
Q^*_{1k}&=\int_0^{2\pi} d\phi e^{-j^2D T}\cos(j\phi) Q^*_k (\phi)\\
Q^*_{2k}&=\int_0^{2\pi} d\phi e^{-j^2D T}\sin(j\phi) Q^*_k (\phi)
\end{align}

We can define the vectors ${\bf c}_1$ and ${\bf c}_2$ with elements
$c_{1j}$ and $c_{2j}$, respectively, as well as the matrices ${\bf
  A}$, ${\bf B}$, ${\bf C}$, ${\bf D}$, with elements $A_{jk}, B_{jk},
C_{jk}, D_{jk}$, respectively, and the vectors ${\bf q}_1$ and ${\bf
  q}_2$ with elements $Q^*_{1j}$ and $Q^*_{2j}$, respectively. This
allows us to define the vectors ${\bf c}^T \equiv ({\bf c}_1^T : {\bf
  c}_2^T)$ and ${\bf q}^T \equiv ({\bf q}_1^T : {\bf q}_2^T)$, where
$T$ denotes the transpose, and the matrix
\begin{align}
{\bf M} = \begin{pmatrix}
{\bf A} & {\bf B}\\
{\bf C} & {\bf D}
\end{pmatrix}.
\end{align}
We can then rewrite \erefstwo{c1}{c2} as
\begin{align}
{\bf c} = {\bf M} {\bf c} + {\bf q},
\end{align}
which has as its solution
\begin{align}
{\bf c} &= \left({\bf I}-{\bf M}\right)^{-1} {\bf q},
\end{align}
with ${\bf I}$ the identity matrix. With the coefficients $c_{1j}$ and
$c_{2j}$ thus found, $f(\phi)$ can be obtained from \eref{app:f},
yielding, finally, the steady-state solution $p_{\rm ss} (\phi) =
1/(2\pi) + f(\phi)$.

\section{Mutual information as a function of $\epsP$ and $\epsM$}
\label{sec:EpsPepsM}
\fref{EpsPoEpsM} addresses how the mutual information depends on
$\epsP$ and $\epsM$. To this end, the parameters are varied as $\epsP
= (1-\alpha) \epsilon$ and $\epsM = \alpha \epsilon$; varying $\alpha$
thus keeps the total absolute coupling strength $\epsilon$ constant. The figure
shows that the mutual information is rather insensitive to the relative
values of $\epsP$ and $\epsM$.

\begin{figure*}[t]
\centering
\includegraphics[width=16cm]{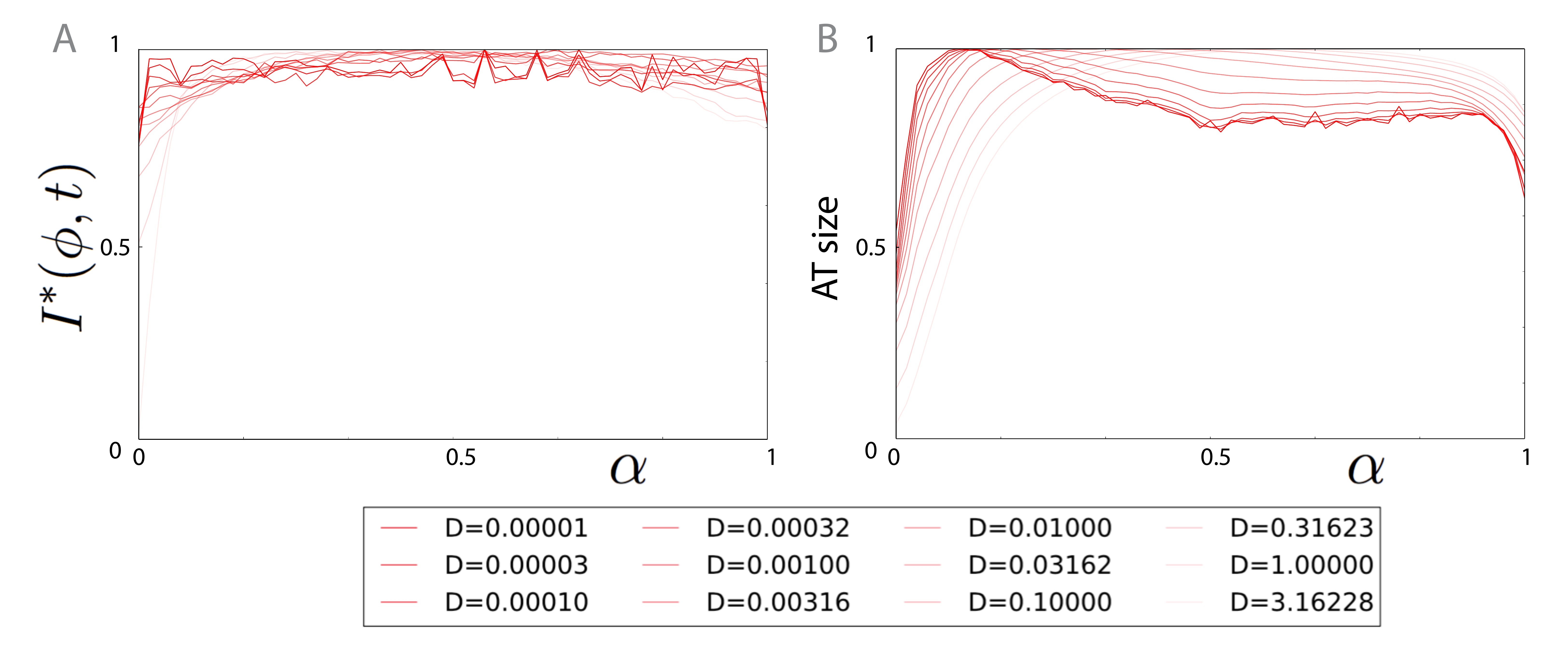}
\caption{Information transmission is not much affected by the relative
  magnitudes of $\epsP$ and $\epsM$ in the coupling function $Z(\phi)$
  (see \fref{cartoon}). We vary $\epsP$ and $\epsM$ via a parameter
  $\alpha$, defined as $\epsP = (1-\alpha) \epsilon$ and $\epsM =
  \alpha \epsilon$; varying $\alpha$ thus keeps the total absolute
  coupling strength constant. We vary $\alpha$ and the diffusion
  constant $D$, and optimize over $\epsilon$ and $\omega_0$, keeping
  $\Delta \phi_{12}=\Delta \phi_{23}=\pi/2$ constant in all
  simulations.  (A) The maximal mutual information $I^*(\phi,t)$,
  obtained by optimizing $I(\phi,t)$ over $\epsilon$ and $\omega_0$,
  as a function of $\alpha$, for different values of $D$. It is seen
  that for most values of $D$, $I^*(\phi;t)$ is quite independent of
  $\alpha$. (B) The size of the Arnold Tongue of the stochastic system
  as a function of $\alpha$, for different values of $D$. The size is
  defined as $\int_{\omega_0^{\rm min}}^{\omega_0^{\rm max}}d\omega_0
  \int_{\epsilon^{\rm min}}^{\epsilon^{\rm max}}d\epsilon I(\phi;t) /
  I^*(\phi;t)$, with $\omega_0^{\rm min}/\omega = 0.4$, $\omega_0^{\rm
    max}=2.7$, $\epsilon_{\rm min}/\omega = 0$, $\epsilon_{\rm
    max}/\omega=5$. It is seen that, except for the low and high
  values of $\alpha$, the size of the Arnold Tongue of the stochastic
  system is fairly independent of $\alpha$.  \flabel{EpsPoEpsM}}
\end{figure*}


\end{document}